\documentclass[12pt]{article}%
\usepackage[nosort]{cite}
\usepackage{graphicx}
\usepackage{multicol}
\usepackage{amsfonts}
\usepackage{amsmath}
\usepackage{amssymb}
\usepackage{amsthm}
\usepackage{heck}
\usepackage{afterpage}
\usepackage{setspace}
\usepackage{verbatim}
\usepackage{color}
\usepackage{longtable}
\usepackage{float}
\usepackage{subcaption}
\usepackage{epsfig}
\usepackage{epstopdf}
\usepackage{adjustbox}
\usepackage{tikz}
\usepackage[margin=1in]{geometry}
\usepackage{titletoc}
\usepackage{hyperref}%
\usepackage{mathrsfs}
\usepackage{dsfont}
\usepackage{algorithm}
\usepackage{algpseudocode}
\usepackage{pgfplots}
\usepackage{algorithmicx, algpseudocode,algorithm}
\usepackage{caption}
\usepackage{float}
\usepackage{makecell}
\usepackage{mathtools}
\usepackage{pdflscape}
\usepackage{array}
\usepackage{longtable}
\usepackage{comment}
\usepackage{mystuff}

\numberwithin{equation}{section}
\newsavebox{\mysavebox}

\hypersetup{colorlinks,citecolor=black,filecolor=black,linkcolor=black,urlcolor=black}

\usepackage[capitalize, noabbrev]{cleveref}
\usetikzlibrary{decorations.markings}
\usetikzlibrary{decorations.markings}
\usetikzlibrary{hobby}
\usepackage{tikz}
\usepackage{tikz-cd}
\usetikzlibrary{arrows}
\usetikzlibrary{arrows.meta}
\usetikzlibrary{arrows,decorations.pathmorphing}
\usetikzlibrary{shapes.geometric,calc,arrows, positioning,shapes.misc,decorations.markings}
\tikzset{
  big arrow/.style={
    decoration={markings,mark=at position 1 with {\arrow[scale=2,#1]{>}}},
    postaction={decorate},
    shorten >=0.4pt},
  big arrow/.default=black}

\pgfdeclarelayer{edgelayer}
\pgfdeclarelayer{nodelayer}
\pgfsetlayers{edgelayer,nodelayer,main}
\pgfplotsset{compat=1.16}
\tikzstyle{none}=[inner sep=0pt]

\theoremstyle{definition}

\crefname{thm}{Theorem}{Theorems}
\crefname{prop}{Proposition}{Propositions}
\crefname{defn}{Definition}{Definitions}
\crefname{lem}{Lemma}{Lemmas}

\tikzstyle{NodeCross}=[draw, shape=circle, cross out, inner sep=0pt, minimum size=6pt,line width=0.25mm]
\tikzstyle{Circle}=[draw, shape=circle, black, fill=black, inner sep=0pt, minimum size=6pt]
\tikzstyle{circle}=[draw, shape=circle, black, fill=black, inner sep=0pt, minimum size=16pt]
\tikzstyle{Star}=[draw, shape=star, fill=black, star points=8, inner sep=0pt, minimum size=8pt]
\tikzstyle{CircleRed}=[draw, shape=circle, black, fill=red, inner sep=0pt, minimum size=6pt]
\tikzstyle{StarP}=[draw={rgb,255: red,128; green,0; blue,128}, shape=star, fill={rgb,256: red,128; green,0; blue,128}, star points=8, inner sep=0pt, minimum size=12pt]
\tikzstyle{ShadedCircRed}=[draw=red, shape=circle, fill={rgb, 255: red,255; green,114; blue, 118}, inner sep=0pt, minimum size=80pt, line width=0.5mm, fill opacity=0.2]
\tikzstyle{ShadedCircRed2}=[draw=red, shape=circle, fill={rgb, 255: red,255; green,114; blue, 118}, inner sep=0pt, minimum size=10pt]
\tikzstyle{ShadedCircRed3}=[draw=black, shape=rectangle, fill={rgb, 255: red,255; green,114; blue, 118}, inner sep=0pt, minimum size=113pt, line width=0.25mm]
\tikzstyle{ShadedCirc}=[draw=red, shape=circle, fill=white, inner sep=0pt, minimum size=45pt,  fill opacity=1.0,  line width=0.5mm]
\tikzstyle{CircleBlue}=[draw, shape=circle, fill=blue, inner sep=0pt, minimum size=6pt]
\tikzstyle{BigCirclePurple}=[draw, shape=circle, fill={rgb,255: red,191; green,0; blue,191}, inner sep=0pt, minimum size=12pt]
\tikzstyle{CirclePurple}=[draw, shape=circle, fill={rgb,255: red,191; green,0; blue,191}, inner sep=0pt, minimum size=5pt]
\tikzstyle{EmptyCircle}=[draw, shape=circle, inner sep=0pt, minimum size=4pt]
\tikzstyle{GreenCircle}=[draw, shape=circle,  fill={rgb,255: red,80; green,200; blue,120}, inner sep=0pt, minimum size=8pt]
\tikzstyle{BrownCircle}=[draw, shape=circle,  fill={rgb,255: red,210; green,105; blue,30}, inner sep=0pt, minimum size=8pt]
\tikzstyle{CirclePurpleSmall}=[draw, shape=circle, fill={rgb,255: red,191; green,0; blue,191}, inner sep=0pt, minimum size=4pt]
\tikzstyle{BigCircleGreen}=[draw, shape=circle, fill={rgb,255: red,0; green,191; blue,0}, inner sep=0pt, minimum size=12pt]
\tikzstyle{BigCircleBlue}=[draw, shape=circle, fill={rgb,255: red,0; green,0; blue,191}, inner sep=0pt, minimum size=12pt]
\tikzstyle{BigCircleRed}=[draw, shape=circle, fill={rgb,255: red,191; green,0; blue,0}, inner sep=0pt, minimum size=12pt]
\tikzstyle{BrownCircleSmall}=[draw, shape=circle,  fill={rgb,255: red,210; green,105; blue,30}, inner sep=0pt, minimum size=6pt]
\tikzstyle{SmallCircleBrown}=[draw, shape=circle,  fill={rgb,255: red,210; green,105; blue,30}, inner sep=0pt, minimum size=5pt]
\tikzstyle{SmallCircleRed}=[draw, shape=circle, fill={rgb,255: red,191; green,0; blue,0}, inner sep=0pt, minimum size=6pt]
\tikzstyle{DashedLine}=[-, densely dashed, line width=0.25mm]
\tikzstyle{DottedLine}=[-, dotted, line width=0.25mm]
\tikzstyle{ThickLine}=[-, line width=0.25mm]
\tikzstyle{ArrowLineRight}=[-, -{Stealth[scale=1.25]}, line width=0.25mm, scale=5]
\tikzstyle{ArrowLineRed}=[-, draw={rgb,255: red,191; green,0; blue,0}, -{Stealth[scale=1.75]}, line width=0.1mm, scale=5]
\tikzstyle{RedLine}=[-, draw={rgb,255: red,191; green,0; blue,0}, fill=none, line width=0.5mm]
\tikzstyle{DashedLineThin}=[-, densely dashed, line width=0.125mm, fill=none, draw=black]
\tikzstyle{DottedRed}=[-, dotted, draw={rgb,255: red,191; green,0; blue,0}, fill=none, line width=0.25mm]
\tikzstyle{DashedRed}=[-, densely dashed, draw={rgb,255: red,191; green,0; blue,0}, fill=none, line width=0.25mm]
\tikzstyle{BlueLine}=[-, draw={rgb,255: red,0; green,0; blue,191}, fill=none, line width=0.5mm]
\tikzstyle{ArrowLineBlue}=[-, draw={rgb,255: red,0; green,0; blue,191}, -{Stealth[scale=1.75]}, line width=0.1mm, scale=5]
\tikzstyle{GreenDoubleArrow}=[<->, draw={rgb,155: red,0; green,255; blue,0},  line width= 0.5mm, scale=5]
\tikzstyle{RedDoubleArrow}=[<->, draw={rgb,255: red,255; green,0; blue,0},  line width= 0.5mm, scale=5]
\tikzstyle{BlueDottedLight}=[-, dotted, draw={rgb,255: red,0; green,0; blue,191}, fill=none, line width=0.3mm]
\tikzstyle{BrownLine}=[-, draw={rgb,255: red,210; green,105; blue,30}, fill=none, line width=0.5mm]
\tikzstyle{DottedRed}=[-, dotted, draw={rgb,255: red,191; green,0; blue,0}, fill=none, dotted, line width=0.5mm]
\tikzstyle{DottedPurple}=[-, dotted, draw={rgb,255: red,191; green,0; blue,191}, fill=none, dotted, line width=0.5mm]
\tikzstyle{BlueDottedLight}=[-, dotted, draw={rgb,255: red,0; green,0; blue,191}, fill=none, line width=0.5mm]
\tikzstyle{ArrowLinePurple}=[-, draw={rgb,255: red,191; green,0; blue,191}, -{Stealth[scale=1.75]}, line width=0.5mm, scale=5]
\tikzstyle{DashedLineGreen}=[-, densely dashed, draw={rgb,255: red,74; green,103; blue,65}, line width=0.25mm]
\tikzstyle{LineGreen}=[-, draw={rgb,255: red, 74; green,200; blue,65}, line width=0.5mm]
\tikzstyle{ArrowLineGreen}=[-, draw={rgb,255: red,0; green,191; blue,0}, -{Stealth[scale=1.75]}, line width=0.5mm, scale=5]
\tikzstyle{GreenLine}=[-, draw={rgb,255: red,0; green,191; blue,0}, fill=none, line width=0.5mm]
\tikzstyle{PurpleLine}=[-, draw={rgb,255: red,191; green,0; blue,191}, fill=none, line width=0.5mm]
\tikzstyle{PPurpleLine}=[-, draw={rgb,255: red,191; green,0; blue,191}, fill=none, line width=2.5mm]
\tikzstyle{DPurpleLine}=[-, dotted, draw={rgb,255: red,191; green,0; blue,191}, fill=none, line width=0.5mm]
\tikzstyle{SBrownLine}=[-, draw={rgb,255: red,191; green,0; blue,191}, fill=none, opacity=0.35, line width=2.5mm]
\tikzstyle{DottedBlue}=[-, dotted, draw=blue, fill=none, dotted, line width=0.5mm]
\tikzstyle{DashedPurpleLine}=[-, densely dashed, draw={rgb,255: red,191; green,0; blue,191}, fill=none, line width=0.5mm]
\tikzstyle{SmallCircleBlue}=[draw, shape=circle, fill=blue, inner sep=0pt, minimum size=5pt]
\tikzstyle{SmallCirclePurple}=[draw, shape=circle, fill={rgb,255: red,191; green,0; blue,191}, inner sep=0pt, minimum size=5pt]
\tikzset{snake it/.style={decorate, decoration=snake}}

\tikzset{
dashstar/.style={
 dash pattern=on 5pt off 5pt,
 postaction={
  decorate,
  decoration={
   markings,
   mark=between positions 9pt and 1 step 10pt with {
     \node[color=red] {*};
   }
  }
 }
},
dashstarstar/.style={ 
 dash pattern=on 5pt off 10pt,
 postaction={
   decorate,
   decoration={
     markings,
     mark=between positions 10pt and 1
          step 15pt
           with {
            \node at (-2pt,0pt) {\pgfuseplotmark{star}};
            \node at (2pt,0pt) {\pgfuseplotmark{star}};
           }
   }
 }
}
}

\begin{document}

\date{January 2026}

\title{The Topological Equivalence Principle: \\[4mm] On Decoupling TFTs from Gravity}

\institution{PENN}{\centerline{$^{1}$Department of Physics and Astronomy, University of Pennsylvania, Philadelphia, PA 19104, USA}}
\institution{KITP}{\centerline{$^{2}$Kavli Institute for Theoretical Physics,
University of California
Santa Barbara, CA 93106 USA}}
\institution{PENNmath}{\centerline{$^{3}$Department of Mathematics, University of Pennsylvania, Philadelphia, PA 19104, USA}}

\authors{
Charlie Cummings\worksat{\PENN,\KITP}\footnote{e-mail: \texttt{charlie5@sas.upenn.edu}} and
Jonathan J. Heckman\worksat{\PENN,\PENNmath}\footnote{e-mail: \texttt{jheckman@sas.upenn.edu}}
}

\abstract{
Topological field theories (TFTs) play an important role in characterizing the deep infrared (IR) of many quantum systems with a mass gap, as well as the global symmetries of quantum field theories (QFTs) decoupled from gravity. In gravitational asymptotically AdS spacetimes, TFT sectors which are putatively decoupled from local metric data are nevertheless non-perturbatively sensitive to Newton's constant via a sum over topologically distinct saddle point configurations. Tracking the fate of this non-decoupling in the boundary dual, we argue that in spite of appearances, this dependence on Newton's constant extends to local metric fluctuations. Said differently, TFTs are in the Swampland. In tandem with earlier results on the absence of global symmetries in theories with subregion-subregion duality, this also establishes that topological operators of boundary systems with a gravity dual are always non-topological in the bulk.}

\maketitle

\enlargethispage{\baselineskip}

\setcounter{tocdepth}{2}


\newpage

\section{Introduction}

Topological field theories play an important role in many physical settings, including the deep infrared (IR) of systems with a mass gap \cite{Witten:1988hf,Witten:1988ze,Wen:1991sca, Kitaev:2005hzj}. Such structures have also recently figured prominently in the modern formulation of generalized global symmetries in quantum field theories (QFTs) decoupled from gravity \cite{Gaiotto:2014kfa}. In that context, topological symmetry operators can support a non-trivial TFT on their worldvolume, leading to rich topological structures in many interacting QFTs.

In quantum gravity, however, there is a general lore that topology is a mutable construct. A related point is that one does not expect any global symmetries.\footnote{See e.g., references \cite{Misner:1957mt, Banks:1988yz, Susskind:1995da, Polchinski:2003bq, Banks:2010zn, Harlow:2018tng, Rudelius:2020orz, Heckman:2024oot, Bah:2024ucp}.} Precisely because TFTs are independent of local metric data,\footnote{Namely, local fluctuations of the metric. One of course can still entertain topological invariants such as the Euler characteristic and Pontryagin classes. Formulations of low-dimensional gravity itself in terms of a quasi-topological system (see e.g., \cite{Witten:1988hc, Witten:2007kt, Collier:2023fwi}) are of course interesting, but are also \textit{maximally} coupled to gravity. Here we seek TFTs which are formulated independent of Newton's constant.} it is natural to ask what role such systems play in gravity. Indeed, while there are many examples of topological interaction terms, most known examples also involve more standard metric dependent terms in an effective action, including, e.g., kinetic terms and higher derivative couplings.

In this note we ask whether TFTs can truly remain decoupled from gravity, i.e., whether they are completely independent of Newton's constant.\footnote{For earlier discussions in a similar vein, see in particular reference \cite{Rudelius:2020orz}. A related comment is that if one assumes the completeness of the spectrum of states in quantum gravity \cite{Polchinski:2003bq}, one expects only invertible TFTs (i.e., those with a one-dimensional Hilbert space) to possibly persist \cite{Montero:2020icj}. Our analysis will not assume completeness and applies equally well to invertible as well as non-invertible TFTs. We thank M. Montero for a discussion on this point.} In what follows, we confine our attention to gravitational spacetimes in three or more spacetime dimensions.

On general grounds one might suspect that one cannot fully decouple TFTs from gravity since the gravitational path integral involves a sum over many different spacetime backgrounds, whereas a TFT (viewed as a special example of a QFT) is typically formulated on a fixed background.\footnote{Perhaps more pedantically one might wish to distinguish TFTs from TQFTs, but we shall always mean TFTs which make sense as quantum mechanical systems.} Letting $S_{\mathrm{grav}}$ denote the action for the gravitational sector, i.e., the metric and any matter which sources stress energy, and letting $S_{\mathrm{TFT}}$ denote the action for the decoupled TFT, the full path integral (in Euclidean signature) takes the schematic form:
\begin{equation}
Z^{\mathrm{full}} = \int [D \Phi_{\mathrm{grav}}] [D \Phi_{\mathrm{TFT}}] \exp^{-S_{\mathrm{grav}}[\Phi_{\mathrm{grav}}] - S_{\mathrm{TFT}}[\Phi_{\mathrm{TFT}}]}.
\end{equation}
Summing over all gravitational saddle point configurations results in an approximation in terms of a weighted sum over different spacetimes $M$, whose weights $w_M$ depend on the details of the gravitational sector:\footnote{Making this intuition precise can be somewhat challenging because one often does not have full access to the necessary saddle point configurations. For example, one might speculate that non-trivial cancellations between saddle point configurations could occur.}
\begin{equation} \label{eqn:ZFULL}
Z^{\mathrm{full}} \simeq \underset{M}{\sum} w_{M} Z^{\mathrm{TFT}}[M].
\end{equation}
So at this level, there is a sense in which correlators of the TFT still depend non-perturbatively on the value of Newton's constant. In other words, even though the decoupled TFT is insensitive to the local data of the metric, the TFT fields are constrained to propagate on the same topological manifold as the gravitational degrees of freedom. We call this the ``topological equivalence principle,'' as the usual equivalence principle can be formulated as saying that all fields couple to the same local metric data.

We emphasize that line \eqref{eqn:ZFULL} is \emph{not} simply a one-loop approximation of the full path integral, it is the exact path integral $Z^{\mathrm{full}}$. What is special about the TFT being a \emph{topological} field theory is that the fixed-$M$ factorization of the gravitational and TFT fields in \eqref{eqn:ZFULL} holds exactly. If $Z^{\mathrm{TFT}}[M]$ was the partition function of a non-topological quantum field theory, then the fixed-$M$ factorization of the gravitational and TFT fields would not hold beyond the one-loop approximation, as interactions between the gravitational fields and TFT fields would lead to a more complicated expansion in $M$.

As written, the relation of line \eqref{eqn:ZFULL} is somewhat heuristic since it requires us to adequately approximate the gravitational path integral. Moreover, even if we can directly carry out such a sum, it is natural to ask whether the mixing between the TFT sector and gravity only occurs non-perturbatively. Our aim will be to sharpen these statements by considering the case of bulk gravitational systems on asymptotically AdS spacetimes. In this case, one expects a large $N$ CFT dual description, and the presence of a putative decoupled TFT sector means that the boundary theory Hilbert space factors as $\mathcal{H}_{\mathrm{CFT}} \otimes \mathcal{H}_{\mathrm{edge}}$, i.e., the large $N$ degrees of freedom necessary to construct a gravity dual arise from $\mathcal{H}_{\mathrm{CFT}}$, while $\mathcal{H}_{\mathrm{edge}}$ serves to characterize possible edge mode degrees of freedom
associated with the bulk TFT.\footnote{The edge mode theory could also be topological.}

Such a factorization must hold, as follows from the fact that operators of the TFT $\widetilde{\mathcal{U}}$ are assumed to be independent of local metric fluctuations in the first place.\footnote{Note that this does not mean the extra sector characterized by $\mathcal{H}_{\mathrm{edge}}$ needs to be a TFT. For example, a priori, the bulk TFT might be an abelian Chern-Simons theory and the boundary system could be a chiral boson.} Indeed, it is helpful to note that for topological symmetry operators of the dual CFT which act non-trivially on (i.e., topologically link / intersect)  interacting degrees of freedom of the CFT, the argument of \cite{Heckman:2024oot} establishes that any candidate bulk operator $\widetilde{\mathcal{U}}$ can be interpreted as the insertion of a dynamical brane in the bulk with non-zero tension, i.e., this object couples to local metric fluctuations.\footnote{For example, in the case of a D-brane, $\mathcal{U}$ is the path integral for the WZ terms of the D-brane, and $\widetilde{\mathcal{U}}$ is the path integral for the D-brane, including the metric dependent DBI action terms.} This imposes strong constraints on possible bulk TFT sectors, but does not by itself exclude the presence of such sectors. Indeed, letting $\mathcal{A}_{\mathrm{CFT}}$ denote the operator algebra which acts on $\mathcal{H}_{\mathrm{CFT}}$, this shows that any putative topological $\widetilde{\mathcal{U}}$ dual to an operator $\mathcal{U}$ in the CFT algebra must commute with all of $\mathcal{A}_{\mathrm{CFT}}$, i.e., it is in $\mathcal{Z}_{\mathrm{CFT}}$, the center of $\mathcal{A}_{\mathrm{CFT}}$ (which is trivial within a fixed superselection sector). More generally, we can imagine deforming the CFT by including additional states in the factorized form $\mathcal{H}_{\mathrm{CFT}} \otimes \mathcal{H}_{\mathrm{edge}}$, where $\widetilde{\mathcal{U}}$ acts non-trivially on the states of $\mathcal{H}_{\mathrm{edge}}$.

To fully exclude the existence of a decoupled TFT sector, we proceed via a proof by contradiction, showing that the bulk path integral and boundary system path integral computations disagree when the TFT is assumed to fully decouple from local metric fluctuations. The outline of our argument is as follows. First, we consider the Hilbert space defined by the TFT itself. In this setting, one can view the radial direction of AdS as specifying a time direction, and the boundary theory as constructing normalized states of $ \mathcal{H}_{\mathrm{TFT}}$, the Hilbert space of the TFT. We assume from the outset that in the bulk TFT, there exist both an operator $\widetilde{\mathcal{U}}$ and a state $\vert \Psi \rangle$ such that:\footnote{Otherwise there is a sense in which the TFT is ``completely trivial''. Note also that even if $\dim \mathcal{H}_{\mathrm{TFT}} = 1$, i.e., we have an invertible TFT, this is still a non-trivial statement.}
\begin{equation}
\bra{\Psi}\widetilde{\mathcal{U}}\ket{\Psi}  \neq d_{\widetilde{\mathcal{U}}}.
\end{equation}
Here, $d_{\widetilde{\mathcal{U}}}$ is the quantum dimension of $\widetilde{\mathcal{U}}$.\footnote{If $\widetilde{\mathcal{U}}$ is invertible, then $d_{\widetilde{\mathcal{U}}} = 1$.} Our aim will be to compute $Z^{\mathrm{bulk}}_{\widetilde{\mathcal{U}}}$, the bulk path integral with $\widetilde{\mathcal{U}}$ inserted, and compare this to $Z^{\mathrm{bdry}}_{\mathcal{U}}$, the value computed in the boundary system with a putative boundary dual $\mathcal{U}$ inserted. In particular, we establish that in general:
\begin{equation}\label{eqn:ZnotZ}
Z^{\mathrm{bulk}}_{\widetilde{\mathcal{U}}} \neq Z^{\mathrm{bdry}}_{\mathcal{U}}.
\end{equation}

Our main observation is that in the bulk gravitational path integral, more than one
saddle point configuration can contribute, and the dominant saddle point depends on the ratio of the
AdS length and Planck length $L_{\mathrm{AdS}} / L_{\mathrm{pl}}$,
which is in turn controlled by the number of CFT degrees of freedom in the boundary system. On the other hand, the computation of $Z^{\mathrm{bdry}}_{\mathcal{U}}$ is---by construction---independent of $L_{\mathrm{AdS}} / L_{\mathrm{pl}}$. Assuming
that we have a parametric family of dual CFTs controlled by the ratio $L_{\mathrm{AdS}} / L_{\mathrm{pl}}$, varying this ratio whilst remaining in the semi-classical gravitational regime results in line \eqref{eqn:ZnotZ}.\footnote{If we could not vary this ratio one would not expect to have gravitational observables which depend smoothly on Newton's constant $G_{N}$. This would be a violent departure from the assumed existence of a semi-classical gravity dual. See also \cite{Schlenker:2022dyo, Baume:2023kkf, Liu:2025ikq} for further comments on observables which depend smoothly on $G_{N}$.}

What are the relevant saddle points to consider? Consider the case where $\mathcal{U}$ is a topological codimension-one defect operator (i.e., a zero-form symmetry operator) of the boundary theory acting on the edge mode Hilbert space. It suffices to consider boundary Euclidean spacetimes of the form $S^1 \times M_{D-1}$, i.e., we work on a spatial $(D-1)$-manifold, and the $S^1$ specifies a thermal circle.\footnote{This same geometry was used in \cite{Heckman:2025isn} to determine the tension of the dynamical branes dual to boundary theory topological symmetry operators. In \cite{Heckman:2025isn} it was shown that the thermal expectation value of a symmetry operator in the boundary CFT specifies a change in the bulk theory free energy, thus providing a way to extract the tension of the branes in the bulk.} The two bulk saddle point configurations of interest are thermal AdS and the Schwarzschild-AdS spacetimes \cite{Hawking:1982dh, Witten:1998zw}, and the transition between the dominant saddle depends on the relative value of the inverse temperature $\beta$ versus $L_{\mathrm{AdS}}$, the AdS radius. In the case where $\mathcal{U}$ is instead a $p$-form symmetry operator, similar considerations hold for Euclidean spacetimes of the form $S^{p} \times S^{q}$ with $p + q = D$ since Hawking-Page transitions are expected to exist in such situations as well \cite{Aharony:2019vgs}.

Consequently, there are no ``truly topological'' operators of the bulk since there is always a non-perturbative dependence on Newton's constant. But once we establish this, even more must hold: since the edge mode sector no longer completely decouples from the rest of the large $N$ CFT, the seemingly decoupled $\mathcal{U}$'s must, in fact act on states of the large $N$ CFT. Since this is simply the setup of reference \cite{Heckman:2024oot}, we conclude that the corresponding bulk dual branes must carry a non-zero tension. An important consequence is that in spite of appearances, the fields of the TFT sector directly couple to local fluctuations of the bulk metric, e.g., through standard kinetic terms (as happens in known top down constructions of the AdS/CFT correspondence). As a consequence, there is a sharp sense in which TFTs are in the Swampland \cite{Vafa:2005ui}.\footnote{See e.g., \cite{vanBeest:2021lhn, Agmon:2022thq} for recent reviews of the Swampland program.} In tandem with reference \cite{Heckman:2024oot}, the absence of any decoupled TFT sectors means that there are no candidate global topological symmetry operators in the bulk, i.e., no global symmetries.

The apparent tension between the bulk and boundary computations can be viewed as a simplified avatar of the ``factorization puzzle,'' namely that apparently decoupled QFT sectors may nevertheless be coupled through bulk gravitational dynamics (i.e., wormhole configurations).\footnote{See e.g., \cite{Maldacena:2004rf, Witten:1999xp, Hubeny:2007xt, Schlenker:2022dyo}, and for the role of such configurations in the context of generalized symmetries, see e.g., \cite{Baume:2023kkf, Bah:2022uyz,  Heckman:2025lmw, Torres:2025jcb}.}

We also present a Lorentzian signature version of these statements in the case of zero-form symmetries. In this case, it suffices to consider the thermofield double, as specified by thermal AdS, and the AdS eternal black hole configuration, i.e., a macroscopic Lorentzian wormhole connecting two copies of the original CFT. Acting with the putative $\widetilde{\mathcal{U}}$ on one side of the black hole leads to strong enough constraints to conclude that the identity is the only defect operator in the theory, so the TFT must be trivial.


The rest of this note is organized as follows. In section \ref{sec:REVIEW} we briefly review the argument of \cite{Heckman:2024oot} which establishes that the topological symmetry operators of a large $N$ holographic CFT are dual to dynamical objects in the bulk. This eliminates most candidate global symmetries of the bulk gravity dual but in principle allows one to contemplate a decoupled TFT sector which is ``transparent'' to the gravitational sector. In section \ref{sec:TFT} we formulate our setup for adding a TFT to the bulk, and in section \ref{sec:OBS} we turn to TFT ``observables,'' i.e., expectation values of TFT operators. In section \ref{sec:CONTRADICT} we establish that TFTs are not actually decoupled in Euclidean gravity, and in section \ref{sec:LORENTZ} we turn to a Lorentzian signature version of this argument.
We present our conclusions in section \ref{sec:CONC}. Appendix \ref{app:EXAMPLES} presents some additional details of how the argument works for Chern-Simons and BF theories.

\section{Bulk Duals of Topological Operators} \label{sec:REVIEW}

In this section we briefly review the argument of \cite{Heckman:2024oot} which establishes that the gravity dual of a topological symmetry operator is a dynamical brane. Consider first the case of $p$-form symmetries for $p \geq 0$. The main assumption of \cite{Heckman:2024oot} is that we have a candidate topological symmetry operator $\mathcal{U}$ which acts non-trivially on an operator $\mathcal{O}$ of the CFT. There is a bulk dual object $\widetilde{\mathcal{U}}$ since there is a bulk symmetry topological field theory / symmetry theory sitting near the conformal boundary (i.e., a ``sliver'') of the full gravity dual.\footnote{See in particular \cite{Witten:1998wy, Aharony:1998qu, Heckman:2024oot}.} Next, consider the linking of $\mathcal{U}$ with $\mathcal{O}$. The bulk dual of $\mathcal{O}_{\mathrm{CFT}}$ is, near the boundary, obtained by applying a suitable smearing kernel \cite{Hamilton:2006az}, resulting in an operator $\widetilde{\mathcal{O}}(z)$, where $z$ denotes the local radial coordinate in the Poincar\'{e} patch. On the other hand, this smearing operation means that the linking between $\mathcal{U}$ and $\mathcal{O}$ amounts to having the symmetry act, resulting in a new operator $\mathcal{O}^{(\mathcal{U})}$. Comparing the bulk dual of the smeared operators $\widetilde{\mathcal{O}^{(\mathcal{U})}}$ and $\widetilde{\mathcal{O}}$, we observe a jump in the bulk stress energy. This apparent jump in the stress energy is sourced by $\widetilde{\mathcal{U}}$, i.e., the bulk object is a dynamical brane which couples to local metric fluctuations. Similar considerations hold for $(-1)$-form symmetries, since in this case we have a domain wall configuration in the bulk, i.e., there is a jump in bulk field profiles as the wall is pulled away from the conformal boundary.

From this perspective, the operator $\mathcal{U}$ should be viewed as defining a topological subsector of the full brane dynamics. This is, for example, what happens with the WZ terms of a D-brane \cite{Green:1996bh}. This is also in accord with the way symmetry operators are realized in top down constructions,\footnote{See e.g., \cite{Heckman:2022muc, GarciaEtxebarria:2022vzq, Apruzzi:2022rei, Heckman:2022xgu, Cvetic:2023plv, Cvetic:2023pgm, Bergman:2024aly, Cvetic:2025kdn, Bah:2025vfu}.} where a ``brane at infinity'' has all of its local dynamics frozen out near the conformal boundary of AdS.

A corollary of this analysis is that it excludes \textit{most} global $p$-form symmetries in the gravitational theory with $p \geq 0$. Indeed, if we had a global symmetry in the bulk, we would have a corresponding topological symmetry operator $\widetilde{\mathcal{U}}$. Via the extrapolate dictionary \cite{Banks:1998dd, Harlow:2011ke}, this defines a topological operator of the boundary CFT. Assuming that this topological operator links with operators of the CFT, we again learn that the bulk dual operator is actually not topological, a contradiction.

There is a potential loophole in the above argument because one could in principle entertain a bulk TFT which is completely decoupled from the gravitational sector. In this case, the candidate $\mathcal{U}$ of the boundary system will act trivially on all operators of the CFT. As such,  the argument of \cite{Heckman:2024oot} would seemingly not apply since the $\mathcal{U}$ constructed in this way has nothing ``to act on''. Our aim in this note will be to address precisely this situation, establishing first of all that no such decoupled TFTs actually exist, and as a corollary, that there are no global symmetries in AdS.

\section{Adding a Decoupled TFT to the Bulk} \label{sec:TFT}

In this section we give a more precision formulation of what it would mean to have a fully decoupled TFT in the context of the AdS/CFT correspondence. Since we will eventually argue that such a setup leads to a contradiction, we shall seek to give as broad a definition as possible to maximize the scope of the resulting statements.

To begin, we assume that we have a $(D+1)$-dimensional gravitational theory in $\mathrm{AdS}_{D+1}$ described by semi-classical gravity, possibly coupled to additional QFT sectors \cite{Maldacena:1997re,Witten:1998qj}.\footnote{It is worth noting that in all known top down constructions of the AdS/CFT correspondence, there is always a non-trivial matter sector, and moreover, it is even challenging to get scale separated AdS rather than a higher-dimensional geometry of the form $\mathrm{AdS} \times X$. See e.g., \cite{Ooguri:2006in, Demirtas:2021nlu, Collins:2022nux} for recent discussions on the existence of scale separated solutions and its relation to the Swampland.} By this, we mean that these QFT sectors have a non-trivial stress energy tensor, i.e., they couple to local fluctuations of the metric. We denote by $S_{\mathrm{grav}}$ the action for this entire sector. We assume that the theory specified by $S_{\mathrm{grav}}$ serves to define a valid AdS/CFT pair, i.e., there is a corresponding CFT dual, with the requisite equivalence between the bulk and boundary path integrals with suitable sources. In particular, the Euclidean signature path integral on $\mathrm{AdS}_{D+1}$ with prescribed boundary conditions
\begin{equation}
Z^{\mathrm{grav}}_{\Phi_{\mathrm{grav}} \sim J} \equiv \underset{\Phi_{\mathrm{grav}} \sim J}{\int} [D \Phi_{\mathrm{grav}}] e^{-S_{\mathrm{grav}}[\Phi_{\mathrm{grav}}]}
\end{equation}
is equivalent \cite{Gubser:1998bc, Witten:1998qj} to the $D$-dimensional $\mathrm{CFT}_{D}$ partition function:
\begin{equation}
Z^{\mathrm{grav}}_{\Phi_{\mathrm{grav}} \sim J} = Z^{\mathrm{CFT}}_{J}.
\end{equation}
We also assume that there is a ``large $N$'' parameter satisfying $N^{\nu} \sim L_{\mathrm{AdS}} / L_{\mathrm{pl}}$ for some $\nu > 0$. While some observables depend on detailed properties of $N$, we assume that since we have a semi-classical gravity dual that there is a suitable notion of smooth observables as obtained by varying the parameter $N$ itself, as in \cite{Schlenker:2022dyo,Liu:2025ikq}.\footnote{It would of course be interesting to find an isolated large $N$ CFT, but we shall not hunt for such a chimera.} Much as in \cite{Heckman:2024oot}, we also assume the bulk gravitational system has spacetime dimension of three or more so that we can entertain a genuine CFT with non-trivial stress energy tensor, as well as black hole backgrounds.\footnote{The case of $\mathrm{AdS}_2 / \mathrm{CFT}_1$ is of course interesting to consider, but in this case certain pathological features also appear such as the appearance of apparent bulk global symmetries.} Summarizing, we assume we have a ``standard'' AdS/CFT setup.

Suppose that we now add a topological field theory (TFT) decoupled from gravity. By this, we mean a TFT which does not depend in any way on the fields appearing in the action $S_{\mathrm{grav}}$, including the metric.\footnote{Note that this excludes topological terms such as the Pontryagin class and Chern-Simons-like terms constructed from the spin connection.} Denote by $S_{\mathrm{TFT}}$ the action for this TFT. A priori, one can entertain adding any reflection positive TFT,\footnote{See \cite{Freed:2016rqq} for a precise definition in the case of invertible TFTs.} leading to an apparent ``infinite landscape'' of AdS/CFT pairs.

That being said, we comment here that some apparent examples in ``standard'' AdS/CFT pairs are actually excluded from the start. For example, on backgrounds such as $\mathrm{AdS}_3 \times S^3$, the isometries of $S^3$ yield $SO(4)$ R-symmetry gauge fields \cite{Maldacena:1997re}. These gauge fields have both a standard Yang-Mills kinetic term, as well as a Chern-Simons term, and the latter dominates at long distances. This sort of term is not really a decoupled TFT for a few reasons. First of all, the kinetic term is still present (even if a subdominant contribution). Additionally, there are many R-charged bulk operators. For all these reasons, such examples are not really decoupled from the bulk gravitational dynamics.

Including the contribution from the TFT sector, the bulk path integral now takes the form:
\begin{equation}
Z^{\mathrm{bulk}}_{\Phi_{\mathrm{grav}} \sim J} \equiv \underset{\Phi \sim J}{\int} [D \Phi_{\mathrm{grav}}] [D \Phi_{\mathrm{TFT}}]e^{-S_{\mathrm{grav}}[\Phi_{\mathrm{grav}}] - S_{\mathrm{TFT}}[\Phi_{\mathrm{TFT}}]}, \label{eqn:bulkZ}
\end{equation}
i.e., we must include suitable boundary conditions for both the gravitational and TFT sectors. To avoid overloading the notation, we have suppressed the boundary conditions for the TFT sector. There are various subtleties in fixing these, but for the most part play no role in what is to follow. The main assumption we make is that we have a genuine operator $\widetilde{\mathcal{U}}$ which can fully detach from the edge mode system at the boundary. Our boundary conditions are implicitly chosen to make this so. We cover this possibility by simply writing $\Phi \sim J$ for the path integral on the righthand side.

Since we expect the TFT to make sense on its own, there is a corresponding edge mode theory, and we refer to the corresponding operator algebra as $\mathcal{A}_{\mathrm{edge}}$. Observe that the edge mode theory for this TFT need not be gapped itself; this is the case, e.g., in many Chern-Simons-like TFTs where the edge mode supports a chiral sector \cite{Witten:1988hf}. On the other hand, it is also possible that the edge mode system is gapped (see e.g., \cite{Belov:2005ze, Belov:2006jd, Kapustin:2010hk}); this can also arise for suitable TFTs and appears prominently, e.g., in the symmetry topological field theory (SymTFT) of QFTs.\footnote{For early work on what has come to be known as the SymTFT, see references \cite{Witten:1998wy, Reshetikhin:1991tc, Turaev:1992hq, Barrett:1993ab,  Fuchs:2002cm, Kirillov:2010nh, Kapustin:2010if, Kitaev:2011dxc, Fuchs:2012dt, Freed:2012bs, Freed:2018cec, Gaiotto:2020iye, Apruzzi:2021nmk, Freed:2022qnc, Kaidi:2022cpf}.}

Because we are assuming (and will subsequently show that this assumption produces a contradiction) that the bulk gravitational and TFT sectors are completely decoupled, the corresponding boundary system Hilbert space and operator algebra must also factorize:
\begin{equation}
\mathcal{H}_{\mathrm{bdry}} = \mathcal{H}_{\mathrm{CFT}} \otimes \mathcal{H}_{\mathrm{edge}} \,\,\, \text{and} \,\,\, \mathcal{A}_{\mathrm{bdry}} = \mathcal{A}_{\mathrm{CFT}} \otimes \mathcal{A}_{\mathrm{edge}}, \label{eqn:Hafactorized}
\end{equation}
in the obvious notation. More generally, if the CFT Hilbert space splits into multiple superselection sectors $\Ha_{\mathrm{CFT}}^{(i)}$, then we could imagine a different edge mode sector $\Ha^{(i)}_{\mathrm{edge}}$ associated to each superselection sector. The total Hilbert space would then instead take the form:
\begin{align}
    \mathcal{H}_{\mathrm{bdry}} = \bigoplus_i\mathcal{H}^{(i)}_{\mathrm{CFT}} \otimes \mathcal{H}^{(i)}_{\mathrm{edge}}\,. \label{eqn:generalfactorization}
\end{align}
The specific factorization of line \eqref{eqn:Hafactorized} can be thought of as a special case of \eqref{eqn:generalfactorization}, where the edge mode Hilbert space $\mathcal{H}^{(i)}_{\mathrm{edge}} = \mathcal{H}_{\mathrm{edge}}$ is the same for all sectors. However, for reasons to be explained shortly, we will not need to entertain this more general form in our analysis. We thus restrict ourselves to the simpler factorization of line \eqref{eqn:Hafactorized} instead.

Consider now a topological operator $\widetilde{\mathcal{U}}$ of the bulk TFT sector.
Via the extrapolate dictionary, this defines a topological operator $\mathcal{U}$
of the boundary system. Since this operator is fully decoupled from the gravitational sector,
it also commutes with all operators in $\mathcal{A}_{\mathrm{CFT}} $.
It can thus be presented as a sum of terms set by central elements:
\begin{equation}\label{eqn:ONTHEEDGE}
\mathcal{U} = \underset{\zeta}{\sum} \zeta \otimes \mathcal{U}^{(\zeta)}_{\mathrm{edge}},
\end{equation}
where the $\zeta$ are in the center of $\mathcal{A}_{\mathrm{CFT}}$, and we have appended a corresponding label $\mathcal{U}_{\mathrm{edge}}^{(\zeta)}$ to topological operators of $\mathcal{A}_{\mathrm{edge}}$.

\section{TFT Correlators} \label{sec:OBS}

We now proceed to the computation of correlators in the TFT sector, i.e., the evaluation of path integrals with TFT operators inserted. There are a few variations on this theme we will need to consider. First of all, we can consider the TFT as a theory in its own right. The main assumption we make is that in backgrounds where the TFT has operators supported on topologically non-trivial cycles, there is at least one non-trivial correlator. Second, we turn to our actual setup in which the bulk consists of a gravitational sector and a decoupled TFT sector. In this case we can use holography to either compute the relevant ``observable'' in the boundary system, or alternatively in terms of bulk quantities. Comparing the two approaches, we observe a mismatch in the answers, signaling a contradiction in our underlying assumptions. We take this contradiction to mean that topological operators of the boundary theory are dual to bulk objects with a tension. The implications of this breakdown in decoupling will eventually lead us to conclude that the fields of the ostensibly decoupled TFT actually couples to local metric fluctuations.

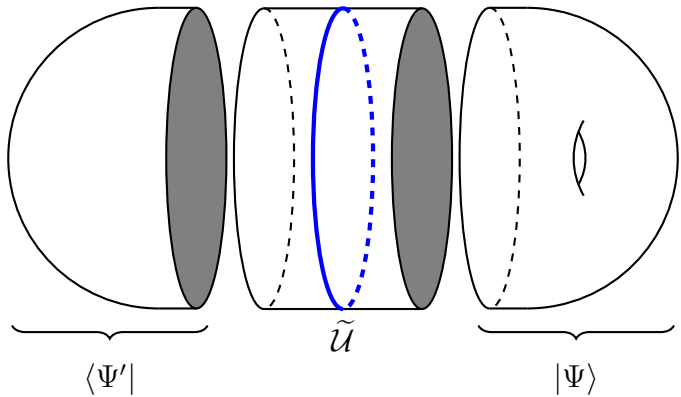
\begin{figure}
    \centering
     \begin{tikzpicture}
    \def\rot{5}
    \def\sep{3}
    \node at (-\sep,0){\begin{tikzpicture}
    \def\hh{0.5}
    \def\xx{-0.75}
    \def\yy{-4.5}

    \draw[thick] (0,-2) arc (90:270:2);
    \draw[thick] (0,-2) -- (\hh,-2);
    \draw[thick] (0,-6) -- (\hh,-6);
    \filldraw[thick,fill=black,fill opacity=0.5] (\hh,-4) ellipse ({2/\rot} and 2);


    \draw [thick,decorate,decoration={brace,amplitude=5pt,mirror,raise=4ex}]
    (-.7-\hh+\xx,-5.5) -- (1.4+\xx,-5.5) node[midway,yshift=-3.5em]{$\bra{\Psi'}$};

    \end{tikzpicture}};

    \node at (0,0.3){\begin{tikzpicture}
    \def\hh{0.5}
    \def\xx{-0.75}
    \def\yy{-4.5}
    \def\ll{0.7}

    \filldraw[thick,fill=black,fill opacity=0.5] (0,0) ellipse ({2/\rot} and 2);

    \draw[thick,xscale={1/\rot}] (-\sep*\rot*\ll,2) arc (90:270:2);
    \draw[thick,dashed,xscale={1/\rot}] (-\sep*\rot*\ll,2) arc (90:-90:2);

    \draw[thick] (0,2) -- (-\sep*\ll,2) (0,-2) -- (-\sep*\ll,-2) ;

    \draw[ultra thick,xscale={1/\rot},blue] (-\sep*\rot*\ll/2,2) arc (90:270:2);
    \draw[ultra thick,dashed,xscale={1/\rot},blue] (-\sep*\rot*\ll/2,2) arc (90:-90:2);

    \node[anchor=north] at (-\sep*\ll/2,-2){$\widetilde{\mathcal{U}}$};

    \end{tikzpicture}};

    \node at (\sep,0){\begin{tikzpicture}
    \def\hh{0.5}
    \def\xx{-0.75}
    \def\yy{-4.5}

    \draw[thick] (0,2) arc (90:-90:2);
    \draw[thick] (0,2) -- (-\hh,2);
    \draw[thick] (0,-2) -- (-\hh,-2);
    \draw[thick,xscale={1/\rot}] (-\hh*\rot,2) arc (90:270:2);
     \draw[thick,dashed,xscale={1/\rot}] (-\hh*\rot,2) arc (90:-90:2);

    \draw[thick] (-.07-\xx,-0.85-\yy-4) arc[start angle=-30, end angle=30, radius=.7cm];
    \draw[thick] (-\xx,-\yy-4) arc[start angle=150, end angle=210, radius=1cm];

    \draw [thick,decorate,decoration={brace,amplitude=5pt,mirror,raise=4ex}]
     (-1.4-\xx,-1.5) -- (.7+\hh-\xx,-1.5) node[midway,yshift=-3.5em]{$\ket{\Psi}$};

    \end{tikzpicture}};

    \end{tikzpicture}
    \caption{Gluing manifolds computes the TFT quantum expectation value $\bra{\Psi'} \widetilde{\mathcal{U}} \ket{\Psi}$.}
    \label{fig:expectationvalue}
\end{figure}

To begin, we consider the TFT in its own right. Placing the TFT on the manifold with boundaries $I \times M_{D}$ for $I = [0,1]$ an interval, we can interpret the interval direction as specifying a time coordinate for the theory. As such, we can prepare states at ``radial time'' $r = 0$ and evolve them with the path integral to radial time $r = 1$.\footnote{We will shortly identify this radial time direction with the radial direction of AdS, hence the terminology.} Consider one such state at radial time $r = 0$, i.e., $\vert \Psi \rangle$. We assume that $M_{D}$ has a topologically non-trivial $m$-cycle as well as the existence of an operator $\widetilde{\mathcal{U}}$ which acts on $\vert \Psi \rangle$ such that:
\begin{equation}\label{eqn:ASSUME}
\bra{\Psi}\widetilde{\mathcal{U}}\ket{\Psi} \neq d_{\widetilde{\mathcal{U}}}
\,,
\end{equation}
with $d_{\widetilde{\mathcal{U}}}$ the quantum dimension of $\widetilde{\mathcal{U}}$. This is a special case of the setup depicted in figure \ref{fig:expectationvalue}. We comment that this is an extremely mild assumption on the structure of any candidate TFT sector.
We are simply asking for one operator and one state which is acted on non-trivially. By a ``trivial TFT,'' we will mean any TFT where we cannot find a $\widetilde{\mathcal{U}}$ and $\vert \Psi \rangle$ satisfying line \eqref{eqn:ASSUME}.

Instead of considering the TFT on a manifold with two boundaries, we can instead consider working on a manifold $M_{D+1}$ with a single boundary $\partial M_{D+1} = M_{D}$. We shall be interested in performing the TFT path integral with the operator $\widetilde{\mathcal{U}}$ inserted:
\begin{equation}
Z^{\mathrm{TFT}}_{\widetilde{\mathcal{U}}} = \int [D \Phi_{\mathrm{TFT}}] \, e^{-S_{\mathrm{TFT}}[\Phi_{\mathrm{TFT}}]} \, \widetilde{\mathcal{U}}.
\end{equation}
The path integral depends on the boundary conditions of the fields $\Phi_{\mathrm{TFT}}$. In the Hilbert space interpretation, this computes an overlap between two different states of the TFT, as specified by the ``capping off'' of the geometry $I \times M_{D}$:
\begin{equation}
Z^{\mathrm{TFT}}_{\widetilde{\mathcal{U}}} =\bra{\Psi'}\widetilde{\mathcal{U}}\ket{\Psi}.
\end{equation}
See figure \ref{fig:expectationvalue} for a depiction.

The end result we get depends on the properties of the state $\vert \Psi^{\prime} \rangle$. This in turn depends on the fate of the $m$-cycle $\Sigma_m$ supporting $\widetilde{\mathcal{U}}$. By abuse of notation we refer to the cycle in $M_{D}$ and its pushforward to $M_{D+1}$ as $\Sigma_m$. Now, it could happen that the non-trivial $m$-cycle $\Sigma_{m}$ can actually become trivial after being continued in the bulk. If it does, then the path integral should actually be ``nearly trivial,'' i.e., if the operator is supported on a topologically trivial cycle:
\begin{equation}
\Sigma_{m} \, \text{trivial in\,} M_{D+1} \, \Rightarrow \frac{Z^{\mathrm{TFT}}_{\widetilde{\mathcal{U}}}}{Z^{\mathrm{TFT}}} = d_{\widetilde{\mathcal{U}}}.
\end{equation}
On the other hand, if $\Sigma_{m}$ remains non-trivial in the bulk $M_{D+1}$, then the path integral will depend on $\widetilde{\mathcal{U}}$:
\begin{equation}
\Sigma_{m} \, \text{non-trivial in\,} M_{D+1} \, \Rightarrow \frac{Z^{\mathrm{TFT}}_{\widetilde{\mathcal{U}}}}{Z^{\mathrm{TFT}}} \neq d_{\widetilde{\mathcal{U}}}.
\end{equation}

Turning next to the bulk path integral where we sum over both the gravitational and TFT sector, we will necessarily encounter different saddle point configurations. These split into two qualitative contributions: those which support a non-trivial $\Sigma_m$ in the bulk, and those which do not. Schematically, however, we can group these contributions much as we did in line \eqref{eqn:ZFULL}:\footnote{Note that this presentation of $Z^{\mathrm{full}}_{\widetilde{\mathcal{U}}}$ generalizes to non-Lagrangian TFTs as well, so our results do not really depend on the TFT being describable by a local action $S_{\mathrm{TFT}}$.}
\begin{equation}
Z^{\mathrm{full}}_{\widetilde{\mathcal{U}}} \simeq \underset{M_{D+1}}{\sum} w_{M_{D+1}} Z^{\mathrm{TFT}}_{\widetilde{\mathcal{U}}}[M_{D+1}].
\end{equation}
Here, $w_{M_{D+1}}$ is a weight that is determined by the details of the gravitational path integral $Z^{\mathrm{grav}}$. If we have at least two topologically distinct saddle point configurations, then our answer will depend on the details of the gravitational sector. This is already a good indication that in spite of appearances, the putative TFT sector still depends on Newton's constant.

Finally, consider the computation with respect to the boundary theory. Here, we need to exercise some care in how we pick our $\widetilde{\mathcal{U}}$ in the first place. Starting with just the TFT sector, there is a corresponding $\mathcal{U}_{\mathrm{edge}} \in \mathcal{A}_{\mathrm{edge}}$ of the corresponding edge mode operator algebra. In principle, the extrapolation in the full bulk system which includes gravity could involve a sum over contributions which are in the center of $\mathcal{A}_{\mathrm{CFT}}$, i.e., we could have (see line \eqref{eqn:ONTHEEDGE}):
\begin{equation}
\mathcal{U}_{\mathrm{full}} = \underset{\zeta}{\sum} \zeta \otimes \mathcal{U}^{(\zeta)}_{\mathrm{edge}}. \label{eqn:Ufull}
\end{equation}
That being said, we can instead consider the special operator given by $\mathbf{id}_{\mathrm{CFT}} \otimes \mathcal{U}_{\mathrm{edge}}$.\footnote{Why not consider the more general class of operators defined by $\mathcal{U}_{\mathrm{full}}$? In the boundary system the partition function will then descend to a weighted sum over different superselection sectors:
\begin{equation}
Z^{\mathrm{bdry}}_{\mathcal{U}} = \underset{\zeta}{\sum} Z^{\mathrm{CFT}}_{\zeta} Z^{\mathrm{edge}}_{\mathcal{U}^{(\zeta)}_{\mathrm{edge}}}.
\end{equation}
Since each $Z^{\mathrm{CFT}}_{\zeta}$ factor can in principle carry a different weight, this can lead to some non-trivial classical dependence on $N$. In the gravity dual, this dependence on the center of $\mathcal{A}_{\mathrm{CFT}}$ can be interpreted as the contribution from possible baby universe sectors \cite{Marolf:2020xie,Gesteau:2020wrk,McNamara:2020uza}. This sort of issue of course hinges on the presence (or absence) of a center of $\mathcal{A}_{\mathrm{CFT}}$, as well as whether the dimension of the baby universe Hilbert space is one-dimensional or high-dimensional. We need not wade into such theological ruminations to establish the main contours of our analysis.}
The main point is that the corresponding operator in the ``gravity + TFT'' bulk already has the desired property of line \eqref{eqn:ASSUME}, i.e., the bulk dual operator acts non-trivially. Furthermore, such operators form a basis for the more general case of \eqref{eqn:Ufull}, so excluding this special case is sufficient to exclude the more general case.\footnote{Central CFT operators, i.e. operators on the baby universe Hilbert space, are not excluded. The reason is that $\mathcal{U}^{(\zeta)}_{\mathrm{edge}} = \mathbf{id}_{\mathrm{edge}}$ is always allowed.} This is also why it is sufficient to consider the specific Hilbert space factorization in line \eqref{eqn:Hafactorized}, as opposed to the more general factorization pattern of line \eqref{eqn:generalfactorization}. Finally, we note that if the CFT algebra $\mathcal{A}_{\mathrm{CFT}}$ has a trivial center (which is expected to be true from Swampland considerations \cite{McNamara:2020uza}), then this special operator is actually completely general.

Let us denote by $Z^{\mathrm{bdry}}_{\mathcal{U}}$ the boundary sector path integral computed in this way. By considering operators of the form $\mathbf{id}_{\mathrm{CFT}} \otimes \mathcal{U}_{\mathrm{edge}}$ within a single superselection sector, and systematically proceeding sector by sector, we can rule out the existence of the bulk TFT in its entirety.

Now, by construction, the boundary sector path integral does not depend on the large $N$ CFT sector at all. As a consequence the evaluation of path integrals factorizes as:
\begin{equation}
Z^{\mathrm{bdry}}_{\mathcal{U}} = Z^{\mathrm{CFT}} Z^{\mathrm{edge}}_{\mathcal{U}}.
\end{equation}
In particular, the expectation value is completely independent of $N$:
\begin{equation}\label{eq:Uvev}
\langle \mathcal{U} \rangle_{\mathrm{bdry}} = \frac{Z^{\mathrm{edge}}_{\mathcal{U}}}{Z^{\mathrm{edge}}},
\end{equation}
in the obvious notation.

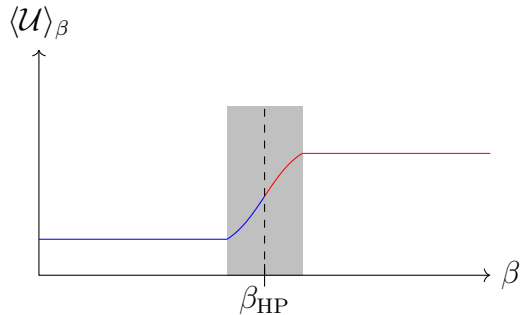
\begin{figure}[t!]
    \centering
    \begin{tikzpicture}[scale=3]
        \def\eps{0.05}
        \def\vv{0.1}
        \def\hh{6}
        \fill[fill opacity=0.25] (1-1/\hh,0) -- (1+1/\hh,0) -- (1+1/\hh,0.75) -- (1-1/\hh,0.75) -- cycle;
        \draw[<->] (0,1) -- (0,0) -- (2,0);
        \node[anchor=south] at (0,1) {$\langle \mathcal{U} \rangle_\beta$};
        \node[anchor=west] at (2,0) {$\beta$};

        \draw[blue] (0,{(1-tanh(1))/4 + \vv}) -- ({1-1/\hh},{(1-tanh(1))/4 + \vv});
        \draw[blue] plot[smooth,domain=-1/\hh:0]({\x+1},{(1+tanh(\x*\hh))/4 + \vv});
        \draw[red] plot[smooth,domain=0:1/\hh]({\x+1},{(1+tanh(\x*\hh))/4 + \vv});
        \draw[red] ({1+1/\hh},{(1+tanh(1))/4 + \vv}) -- ({2},{(1+tanh(1))/4 +\vv});

        \draw[dashed] (1,0) -- (1,0.75);
        \draw (1,-\eps) -- (1,\eps);
        \node[anchor=north] at (1,0) {$\beta_{\mathrm{HP}}$};
    \end{tikzpicture}
    \caption{A graph of $\langle \mathcal{U} \rangle_{\beta}$ as a function of $\beta$, as computed in the bulk dual with fixed boundary conditions. The AdS-Schwarzschild phase is in blue, and the thermal AdS phase is in red. Generically, $\langle \mathcal{U} \rangle_\beta$ is constant in each phase, and the phase transition occurs at the inverse Hawking-Page temperature $\beta_{\mathrm{HP}}$. On the other hand, in the CFT dual, this correlator is controlled by an edge mode system which does not depend on $N$, i.e., the correlation function is expected to be the same on both backgrounds. This signals a contradiction.}
    \label{fig:step}
\end{figure}

It is here that we encounter a clash between our different ways of computing path integrals / partition functions.
On the one hand, we expect on general grounds that the bulk and boundary partition functions must match:
\begin{equation}\label{eqn:NOJOY}
Z^{\mathrm{bulk}}_{\widetilde{\mathcal{U}}} \overset{?}{=} Z^{\mathrm{bdry}}_{\mathcal{U}}.
\end{equation}
On the other hand, the lefthand side involves a sum over different saddle point configurations, and potentially
non-trivial $N$ dependence, while the expectation value of line (\ref{eq:Uvev}) is independent of $N$, i.e., the righthand side seems to always factorize as $Z^{\mathrm{bdry}}_{\mathcal{U}} = Z^{\mathrm{CFT}} Z^{\mathrm{edge}}_{\mathcal{U}}$.

To completely exhibit a contradiction, we next show that varying the ratio $L_{\mathrm{AdS}} / L_{\mathrm{pl}} \sim N^{\nu}$ can change the topology of the bulk saddle. This will establish that line \eqref{eqn:NOJOY} cannot hold, and we take this to mean that there are no decoupled TFT sectors.

\section{Euclidean Saddles and Topology} \label{sec:CONTRADICT}

In the previous section we presented two seemingly related ways to calculate the path integral with $\widetilde{\mathcal{U}}$ inserted.
One route involves working in the boundary theory and the insertion of an operator $\mathbf{id}_{\mathrm{CFT}} \otimes \mathcal{U}_{\mathrm{edge}}$, where the resulting path integral is manifestly independent of $N$. On the other hand, the gravity dual apparently requires a sum over gravitational saddle point configurations, and in turn clearly depends on $N$ through the ratio $L_{\mathrm{AdS}}/ L_{\mathrm{pl}} \sim N^{\nu}$.

In this section we turn to some explicit backgrounds where we can track the dominant saddle point configurations. The main point is that these different saddle point configurations have different bulk topology, and as such contribute differently to the TFT path integral. In particular, this will establish that the TFT cannot be fully decoupled from gravity.

Topologically, we are interested in bulk geometries $M_{D+1}$ which have boundary $\partial M_{D+1} = M_{D}$, and which also retain suitable bundle structures as we approach the boundary. Starting from the boundary geometry, the question becomes how many different ways we can ``fill in the bulk.'' Now, in the case of $M_{D} = S^{1}_{\beta} \times S^{D-1}$, i.e., the boundary system on a thermal circle, there are two dominant saddle point configurations which depend on the size of $\beta$ relative to $L_{\mathrm{AdS}}$.\footnote{These are the saddle point configurations which famously specify the Hawking-Page transition \cite{Hawking:1982dh}. See \cite{Witten:1998zw} for a holographic interpretation.} These two saddles are thermal AdS and AdS-Schwarzschild. In the case of thermal AdS, we have a $S^{1} \rightarrow \mathbb{B}^D$ fibration, where $\mathbb{B}^D$ is a $D$-dimensional ball with boundary $S^{D-1}$. In the case of AdS-Schwarzschild, we instead have a $S^{D-1} \rightarrow \mathbb{B}^{2}$ fibration, where the boundary of the $2$-ball $\mathbb{B}^{2}$ is the thermal circle $S^{1}_{\beta}$.

By inspection, the thermal AdS geometry supports a non-trivial $S^1$, while the AdS-Schwarzschild geometry supports a non-trivial $S^{D-1}$. We have assumed the existence of a non-trivial operator $\widetilde{\mathcal{U}}$ which wraps one of the non-trivial cycles such that:
\begin{equation}
\frac{Z^{\mathrm{TFT}}_{\widetilde{\mathcal{U}}}[M_{D+1}]}{Z^{\mathrm{TFT}}[M_{D+1}]} \neq d_{\widetilde{\mathcal{U}}}.
\end{equation}
The case of a non-trivial topological operator supported on a $(D-1)$-cycle is interpreted in the boundary system as a codimension-one topological defect i.e., a zero-form symmetry operator. The case of a non-trivial topological operator supported on a $1$-cycle is interpreted as a magnetic dual codimension-$(D-1)$ topological defect, i.e., a $(D-2)$-form symmetry operator. Let us comment here that even though we have couched our discussion in terms of the boundary geometry $S^{1} \times S^{D-1}$, similar considerations hold if we instead consider the boundary geometry $S^{1} \times M_{D-1}$ for $M_{D-1}$ some choice of ``spatial'' $(D-1)$-manifold \cite{Birmingham:1998nr}.

Now consider the value of the boundary expectation value $\langle \mathcal{U} \rangle_\beta$, for a fixed $\beta \sim L_{\mathrm{AdS}}$. In units where the Planck length $L_{\mathrm{pl}}$ is fixed, we can vary the bulk AdS radius by varying $N$, and therefore whether $\beta$ is above or below the inverse temperature of the Hawking-Page transition. Thus, at first glance, the value of the correlation function $\langle \mathcal{U} \rangle_\beta$ depends explicitly on the value of $N$. See figure \ref{fig:step} for a depiction of this transition. This is a contradiction with the non-perturbative decoupling between the CFT and the edge mode sector given in line \eqref{eqn:Ufull}, unless the numerical value of $\langle \mathcal{U} \rangle_\beta$ is independent of $\beta$. That this is true for every $\mathcal{U}$ violates line \eqref{eqn:ASSUME}. Thus, the hypothetical TFT must be trivial.

\begin{figure}[t!]
    \centering
    \begin{tikzpicture}
        \def\vv{0.5}
        \def\hh{3}
        \def\nn{3}
        \def\eps{0.25}

        \draw[thick,xscale=\vv] (0,0) circle (1);
        \draw[thick,yscale=\vv] (-\vv,0) arc (180:360:\vv);
        \draw[thick,dashed,yscale=\vv] (-\vv,0) arc (180:0:\vv);
        \draw[thick,xscale=\hh] (0,1) arc (90:270:1);

        \draw [thick,decorate,decoration={brace,amplitude=5pt,raise=4ex}]  ({0},{1.5+2*\eps +3/\nn }) -- ({0},{1.5+2*\eps}) node[midway,xshift=3.5em] {$S^{1}$};

        \draw [thick,decorate,decoration={brace,amplitude=5pt,raise=4ex}]  ({0},{1}) -- ({0},{-1}) node[midway,xshift=5.5em] {$\mathrm{Cone}(S^{D-1})$};

        \foreach\xx in {0,...,\nn}{
            \def\rr{1.5*(1-\xx/(\nn+4))/\nn}
            \filldraw ({-\hh*\xx/\nn},{sqrt(1-(\xx/\nn)*(\xx/\nn)}) circle (0.05);
            \draw[dotted,thick ] ({-\hh*\xx/\nn},{0.5+\eps+sqrt(1-(\xx/\nn)*(\xx/\nn)}) -- ({-\hh*\xx/\nn},{\eps + sqrt(1-(\xx/\nn)*(\xx/\nn)});
            \draw[thick,xscale=\vv] ({-\hh*\xx/\nn / \vv},{0.5+\rr+2*\eps+sqrt(1-(\xx/\nn)*(\xx/\nn))}) circle ({\rr});
        }
    \end{tikzpicture}
    \caption{Depiction of Euclidean thermal AdS. Topologically, this is an $S^1$ fibered over $\mathrm{Cone}(S^{D-1})$. A zero-form symmetry operator $\mathcal{U}$ wrapping the boundary $S^{D-1}$ can be pushed into the bulk, with dual $\widetilde{\mathcal{U}}$. The bulk cycle supporting $\widetilde{\mathcal{U}}$ is contractible, so the thermal correlation function of $\mathcal{U}$ evaluates to the quantum dimension of $\widetilde{\mathcal{U}}$.}
    \label{fig:thermalAdS}
\bigskip
    \centering
    \begin{tikzpicture}
        \def\vv{0.5}
        \def\hh{3}
        \def\nn{3}
        \def\eps{0.25}

        \draw[thick,xscale=\vv] (0,0) circle (1);
        \draw[thick,xscale=\hh] (0,1) arc (90:270:1);

        \draw [thick,decorate,decoration={brace,amplitude=5pt,raise=4ex}]  ({0},{1.5+2*\eps +3/\nn }) -- ({0},{1.5+2*\eps}) node[midway,xshift=3.5em] {$S^{D-1}$};

        \draw [thick,decorate,decoration={brace,amplitude=5pt,raise=4ex}]  ({0},{1}) -- ({0},{-1}) node[midway,xshift=4.5em] {$\mathrm{Cone}(S^{1})$};

        \foreach\xx in {0,...,\nn}{
            \def\rr{1.5*(1-\xx/(\nn+4))/\nn}
            \filldraw ({-\hh*\xx/\nn},{sqrt(1-(\xx/\nn)*(\xx/\nn)}) circle (0.05);
            \draw[dotted,thick ] ({-\hh*\xx/\nn},{0.5+\eps+sqrt(1-(\xx/\nn)*(\xx/\nn)}) -- ({-\hh*\xx/\nn},{\eps + sqrt(1-(\xx/\nn)*(\xx/\nn)});
            \draw[thick] ({-\hh*\xx/\nn},{0.5+\rr+2*\eps+sqrt(1-(\xx/\nn)*(\xx/\nn))}) circle ({\rr});
            \draw[thick,yscale=\vv] ({-\hh*\xx/\nn - \rr},{(0.5+\rr+2*\eps+sqrt(1-(\xx/\nn)*(\xx/\nn)))/\vv}) arc (180:360:{\rr});
            \draw[thick,dashed,yscale=\vv]({-\hh*\xx/\nn - \rr},{(0.5+\rr+2*\eps+sqrt(1-(\xx/\nn)*(\xx/\nn)))/\vv}) arc (180:0:{\rr});

        }
    \end{tikzpicture}
    \caption{Depiction of the Euclidean AdS-Schwarzschild solution. Topologically, this is an $S^{D-1}$ fibered over a $\mathbb{B}^2$, i.e., a 2-ball (disk). A zero-form symmetry operator $\mathcal{U}$ wrapping the boundary $S^{D-1}$ can be pushed into the bulk. The bulk cycle supporting the bulk dual $\widetilde{\mathcal{U}}$ is not contractible, so the thermal correlation function of $\mathcal{U}$ should generically not equal the quantum dimension of $\widetilde{\mathcal{U}}$.}
    \label{fig:blackhole}
\end{figure}

It is also of interest to consider more general boundary geometries of the form $M_{D} = S^{p} \times S^{q}$ for $p+q = D$. In this case, there are again two canonical ways to ``fill in the bulk,'' as dictated by the fibrations $S^{q} \rightarrow \mathbb{B}^{p+1}$ and $S^{p} \rightarrow \mathbb{B}^{q+1}$. While it is more challenging to build bulk gravitational solutions in these situations, the expectation from large $N$ CFTs is that these saddle point solutions characterize a generalized notion of confinement / deconfinement transitions \cite{Aharony:2019vgs}. Indeed, in this case one can wrap a topological operator over the fiber $S^{q}$ (resp. $S^{p}$), resulting in a codimension-$(D-q)$ (resp. $(D-p)$) topological defect in the boundary theory, i.e., a $(p-1)$-form symmetry (resp. $(q-1)$-form symmetry).

Finally, consider $\widetilde{\mathcal{U}}$ which are codimension-one in the bulk. These correspond to $(-1)$-form symmetries of the boundary edge mode system, i.e., they fill the whole boundary. Excluding this situation is somewhat more subtle, but can be addressed if we consider various two-sided systems, i.e., we take two copies of our proposed boundary system. In this case, there is a gravitational saddle which just involve filling in two copies of ``empty AdS,'' i.e., two copies of $M_{D+1} = \mathbb{B}^{D+1}$. There is another gravitational saddle in which we topologically take a cylinder $I \times S^{D}$ with each side of the interval supporting a boundary system, i.e., a Euclidean wormhole of the sort considered in \cite{Witten:1999xp,Maldacena:2004rf,Marolf:2021kjc}.

One might worry that a fine-tuned TFT might happen to trivialize on both saddles on either side of the (generalized) Hawking-Page transition, but remain non-trivial on other more complicated saddles. For such TFTs, we expect that it is possible to engineer phase transitions between these more complicated saddle points that could be used to prove a similar contradiction, but we leave this for future work.
However, such a TFT would be rather ill-behaved. As explained in Appendix \ref{app:EXAMPLES}, the dimension $p$ defects of such a TFT would have a trivial linking matrix $S_p \propto \mathbf{id}$ with the codimension-$(p+1)$ defects, even in higher dimensions.
For any defect $\mathcal{O}_\ell$, the quantum dimension of the defect (defined as $d_\ell = \langle \mathcal{O}_\ell \rangle_{S^D}$) can be computed to be  $d_\ell = (S_p)_{0\ell}/(S_p)_{00}$ via a surgery argument, similar to the 3D case. Thus, for a TFT which happens to be trivial across the (generalized) Hawking-Page transition, the identity will be the only defect with a non-zero quantum dimension. Such a TFT seems rather pathological, but we leave it to future work to determine the saddle points relevant for ruling out such TFTs.

Summarizing, we have established that the bulk does not support full decoupling between the gravitational and TFT sectors.\footnote{For a $\widetilde{\mathcal{U}}$ corresponding to a candidate non-invertible symmetry, we could also have considered a weaker scenario of a fluctuation of a gravitational soliton \cite{Witten:1985xe,McNamara:2019rup,McNamara:2021cuo, Heckman:2024obe} rather than a change in saddle point. The reason is that in the absence of a soliton, the TFT operator can be deformed freely, but in the presence of the soliton, a network of defects remains. Consistency with the decoupling between gravity and the TFT then requires the absence of the symmetry operator. To rule out invertible symmetries, the full saddle point analysis is required. We thank M. H\"ubner for correspondence on this point.}

\subsection{Interpretation of no Decoupled Sectors}

The presence of non-trivial topologies for saddle point configurations means that any bulk TFT is not truly decoupled from gravity, since topological correlators will still depend on Newton's constant. Our saddle point analysis establishes a non-perturbative mixing.
But even more now follows. In the boundary theory, this large $N$ dependence  signals the absence of factorization for the boundary system, i.e., there are non-trivial correlations between the $\mathcal{A}_{\mathrm{CFT}}$ and $\mathcal{A}_{\mathrm{edge}}$ sectors. There are two situations we need to consider:
\begin{itemize}
\item There exists an operator of the edge mode sector which does not commute with the CFT stress tensor.
\item All operators of the edge mode sector commute with the CFT stress tensor.
\end{itemize}
In both cases, the existence of operators in $\mathcal{A}_{\mathrm{edge}}$ which act non-trivially on $\mathcal{A}_{\mathrm{CFT}}$ will signal that the fields of the TFT actually couple to local fluctuations of the bulk metric.

Suppose first that the stress energy tensor of the CFT sector acts non-trivially on some state(s) of the edge mode sector. In this case, the putative topological operator $\widetilde{\mathcal{U}}$ of the bulk will link in the bulk with defects which carry non-zero tension. As such, the general argument of \cite{Heckman:2024oot} can be applied to establish that $\widetilde{\mathcal{U}}$ actually carries stress energy and couples to local fluctuations of the bulk metric. Of course, this contradicts the initial assumption that we had a truly decoupled TFT. It is worth noting that once we have found one such edge mode operator, more clearly follow from its orbit acting on $\mathcal{A}_{\mathrm{edge}}$.

Now, it could still happen that all the operators of $\mathcal{A}_{\mathrm{edge}}$ commute with the stress tensor of $\mathcal{A}_{\mathrm{CFT}}$. For example, this could occur if the edge mode sector is fully gapped. Observe, however, that we now have additional candidate topological operators from our bulk TFT sector (and their extrapolates to the boundary), but which act non-trivially on the large $N$ CFT sector (since the two sectors are not fully decoupled). But this just means we have some additional candidate topological symmetry operators of the large $N$ CFT; precisely the setup of \cite{Heckman:2024oot}. So, we again conclude that the bulk dual of all of these operators are not topological in the bulk, again a contradiction!

The general conclusion from these considerations is that the fields appearing in our candidate ``decoupled TFT sector'' actually couple at the level of local metric fluctuations, a stronger conclusion. A related comment is that this is precisely the situation encountered in all known top down realizations of holography. For example, the 11D supergravity action contains a dynamical three-form field with kinetic term proportional to $dC_3 \wedge \ast dC_3$ as well as a topological term proportional to $C_3 \wedge G_4 \wedge G_4$ \cite{Cremmer:1978km}. Attempting to discard the kinetic term in the deep infrared is acceptable only if gravity has been completely decoupled, including topology change.
If the candidate TFT sector has no consistent coupling to local metric fluctuations, then we can interpret our results as signaling a mixed anomaly between the symmetry operators of the TFT and the gravitational sector associated with topology change.
This perspective was advocated for in e.g., \cite{Jensen:2017eof,Thorngren:2020yht,Heckman:2024obe}.

Summarizing, although our analysis started by establishing a non-perturbative breakdown of topological decoupling, this leads to the stronger statement that operators of this putative TFT sector actually couple to local metric fluctuations. Returning to the discussion of section \ref{sec:REVIEW}, this eliminates the appearance of candidate $p$-form global symmetries for $p \geq 0$.

\section{Lorentzian Interpretation} \label{sec:LORENTZ}

The argument presented in the previous section established the absence of decoupled TFTs by appealing to the structure of the Euclidean path integral. In this section we turn to a Lorentzian signature analysis. We shall be interested in the structure of the thermal mixed state of the boundary system, and its purification in the thermofield double. We will obtain a contradiction by tracking the structure of the thermofield double state above and below the Hawking-Page transition. We focus on the case of bulk codimension-one operators (i.e., boundary zero-form topological symmetry operators) since in this case the topology of an $S^{D-1}$ directly enters as the horizon of the AdS-Schwarzschild black hole. It would be interesting to consider other black objects in a similar vein (presumably the Lorentzian analog of reference \cite{Aharony:2019vgs}), but we defer that analysis to future work.

\begin{figure}
    \centering
    \begin{tikzpicture}
        \node at (0,0) { \begin{tikzpicture}[scale=3]
        \def\hh{0.6}
        \draw[thick] (0.5,0) -- (0,0) -- (0,1) -- (0.5,1);
        \draw[dashed,thick] (0.5,0) -- (0.5,1);
        \draw[thick] (\hh,0) -- (\hh+0.5,0) -- (\hh+0.5,1) -- (\hh,1);
        \draw[dashed,thick] (\hh,0) -- (\hh,1);
        \draw[blue,thick] (\hh,0.5) -- (\hh+0.5,0.5);
        \draw[blue,thick] (0,0.5) -- (0.5,0.5);
        \node[blue,anchor=west] at (\hh+0.5,0.5) {$\Sigma$};

        \end{tikzpicture}};

        \node at (6,0) { \begin{tikzpicture}[scale=3]
        \def\hh{0.6}
        \draw[thick]  (0,0) -- (0,1) (1,1) -- (1,0);
        \draw[thick,decorate,decoration={zigzag,mirror,segment length=10pt,amplitude=-1.5pt}] (0,1) -- (1,1);
        \draw[thick,decorate,decoration={zigzag,segment length=10pt,amplitude=-1.5pt}] (0,0) -- (1,0);
        \draw[dashed,thick] (0,0) -- (1,1);
        \draw[dashed,thick] (1,0) -- (0,1);
        \draw[blue,thick] (0,0.5) -- (1,0.5);
        \node[blue,anchor=west] at (1,0.5) {$\Sigma$};
        \end{tikzpicture}};

    \end{tikzpicture}

    \caption{Left: Penrose diagram of thermal AdS ($\beta > \beta_{\mathrm{HP}}$). Right: Penrose diagram of AdS-Schwarzschild ($\beta < \beta_{\mathrm{HP}}$). For both spacetimes, a Cauchy slice drawn in blue. The TFD is prepared by Euclidean evolution, and continuation along this Cauchy slice.}
    \label{fig:penrose}
\end{figure}
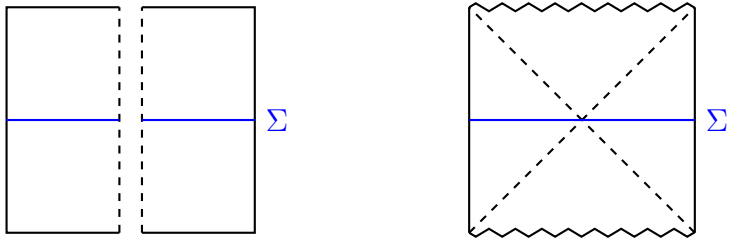

To set conventions, consider first Lorentzian signature $\mathrm{AdS}_{D+1}$ with boundary $\mathbb{R}_{\mathrm{time}} \times S^{D-1}$. On a fixed Cauchy slice, the geometry has the topology of a $\mathbb{B}^D$ with boundary $S^{D-1}$. Consistent dynamics in this spacetime requires a choice of boundary conditions at for all bulk fields, which are to be specified on the asymptotic boundary $\mathbb{R}_{\mathrm{time}} \times S^{D-1}$. For a prescribed set of sources $J$ which determines these boundary conditions, we denote by $\mathcal{H}(\mathbb{B}^D)_J$ the Hilbert space of states for this bulk system. Observe that there are two notions of ``boundary'' now: There is the global AdS time boundary at $t = \pm \infty$, and there is also the radial direction boundary. Here, the Hilbert space makes reference to a fixed time slice, i.e., \textit{not} the radial time direction used in our Euclidean signature discussion.

We shall be interested in the thermal mixed state and its purification to the thermofield double.\footnote{For recent discussions of the SymTFT formalism and its extension to such systems, see references \cite{Heckman:2025lmw,Torres:2025jcb} as well as \cite{Qi:2025tal, Jia:2025bui}.}
Recall that for a system with Hamiltonian $\widehat{H}$ and energy spectrum $E_n$, the state $\vert \mathrm{TFD}\rangle$ is defined as:
\begin{equation}
\vert \mathrm{TFD} \rangle \equiv \frac{1}{\sqrt{Z}} \underset{n}{\sum} e^{-\beta E_{n} / 2} \vert n \rangle \otimes \vert \Theta n \rangle,
\end{equation}
where the presence of $\Theta$ serves as the CRT operator of our two-sided system, and $Z$ denotes the thermal partition function.
We shall be interested in three sorts of TFD states:
\begin{itemize}
  \item $\vert \mathrm{TFD} \rangle_{\mathrm{CFT}}$, the one of the large $N$ CFT
  \item $\vert \mathrm{TFD} \rangle_{\mathrm{edge}}$, the one of the edge mode system
  \item $\vert \mathrm{TFD} \rangle_{\mathrm{bdry}}$, the full boundary system
\end{itemize}
In particular, because we have assumed that the operator algebras of the boundary system completely factorize,
the corresponding thermofield double states satisfy:
\begin{equation}
\vert \mathrm{TFD} \rangle_{\mathrm{bdry}} = \vert \mathrm{TFD} \rangle_{\mathrm{CFT}} \otimes \vert \mathrm{TFD} \rangle_{\mathrm{edge}}.
\end{equation}

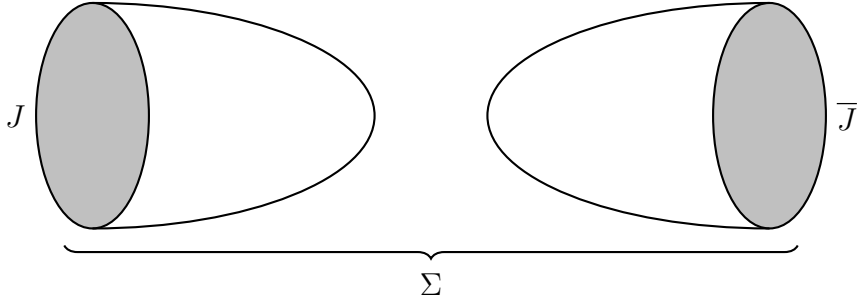
\begin{figure}
    \centering
    \begin{tikzpicture}
        \def\vv{0.5}
        \def\rr{1.5}
        \def\bdrysep{6}
        \def\horsep{1}
        \def\bcsep{-0.5}

        \draw [thick,decorate,decoration={brace,amplitude=5pt,mirror,raise=4ex}] ( {-0.5*\rr*\vv},-1) -- ({\rr*(\bdrysep+0.5*\vv)},-1) node[midway,yshift=-3em]{$\Sigma$};

        \filldraw[fill opacity=0.25,thick,xscale=\vv] (0,0) circle (\rr);
        \draw[thick,xscale={(\bdrysep - \horsep)/(2)}] (0,\rr) arc (90:-90:\rr);

       \filldraw[fill opacity=0.25,thick,xscale=\vv] (\bdrysep*\rr/\vv,0) circle (\rr);
        \draw[thick,xscale={(\bdrysep - \horsep)/(2)}] ({2*\rr*\bdrysep / (\bdrysep - \horsep)},\rr) arc (90:270:\rr);


        \node[anchor=west] at ({(1+\bdrysep+\bcsep)*\rr},0) {$\overline{J}$};


        \node[anchor=east] at ({(-1-\bcsep)*\rr},0) {$J$};

    \end{tikzpicture}
    \caption{A Cauchy slice $\Sigma$ of thermal AdS, along with a manifold-with-boundary preparing boundary conditions $J$ and $\overline{J}$, respectively. We have suppressed the timelike direction. }
    \label{fig:thermalAdSCauchy}
\end{figure}

From the bulk perspective, what Hilbert space is $\ket{\mathrm{TFD}}_{\mathrm{edge}}$ a state of? First, focus on the case $\beta > \beta_{\mathrm{HP}}$, so that we are in the thermal AdS phase. In this case, we essentially have two disconnected copies of AdS, each with a source $J$ and $\overline{J}$. By inspection, then, $ \ket{\mathrm{TFD}}_{\mathrm{edge}}$ is an element of the Hilbert space $\Ha(\mathbb{B}^D)_J \otimes \Ha(\mathbb{B}^D)_{\overline{J}} $ in this phase. See figure \ref{fig:thermalAdSCauchy} for a depiction.

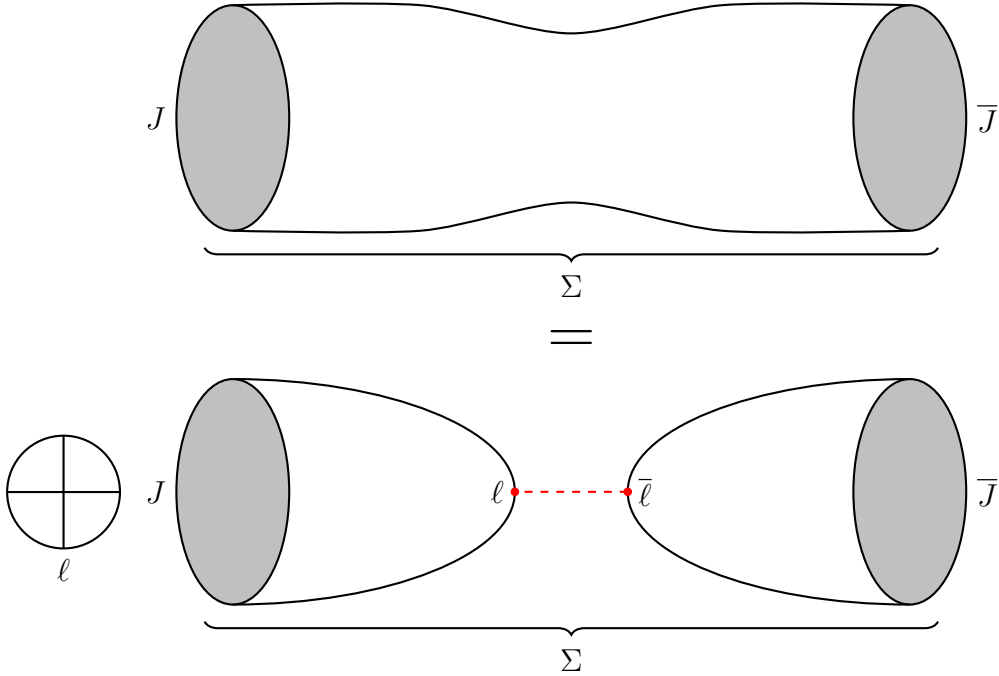
\begin{figure}
    \centering
    \begin{tikzpicture}
        \def\vv{0.5}
        \def\rr{1.5}
        \def\bdrysep{6}
        \def\horsep{1}
        \def\bcsep{-0.5}
        \def\horizon{0.75}
        \def\vert{5}

        \node at (0,0) {\begin{tikzpicture}

        \draw [thick,decorate,decoration={brace,amplitude=5pt,mirror,raise=4ex}] ( {-0.5*\rr*\vv},-1) -- ({\rr*(\bdrysep+0.5*\vv)},-1) node[midway,yshift=-3em]{$\Sigma$};

        \filldraw[fill opacity=0.25,thick,xscale=\vv] (0,0) circle (\rr);
        \filldraw[fill opacity=0.25,thick,xscale=\vv] (\bdrysep*\rr/\vv,0) circle (\rr);
        \draw[thick] plot[smooth,tension=0.5] coordinates {(0,\rr)({\bdrysep*\rr/2 - 2},\rr) ({\bdrysep*\rr/2},\rr*\horizon)
        ({\bdrysep*\rr/2 + 2},\rr)
        (\bdrysep*\rr,\rr)};
        \draw[thick,yscale=-1] plot[smooth,tension=0.5] coordinates {(0,\rr)({\bdrysep*\rr/2 - 2},\rr)({\bdrysep*\rr/2},\rr*\horizon)
        ({\bdrysep*\rr/2 + 2},\rr)
        (\bdrysep*\rr,\rr)};

        \node[anchor=west] at ({(1+\bdrysep+\bcsep)*\rr},0) {$\overline{J}$};


        \node[anchor=east] at ({(-1-\bcsep)*\rr},0) {$J$};
        \end{tikzpicture}};

        \node at (0,-\vert/2) {\Huge$=$};

        \node at (0,-\vert) {\begin{tikzpicture}
        \draw [thick,decorate,decoration={brace,amplitude=5pt,mirror,raise=4ex}] ( {-0.5*\rr*\vv},-1) -- ({\rr*(\bdrysep+0.5*\vv)},-1) node[midway,yshift=-3em]{$\Sigma$};

        \filldraw[fill opacity=0.25,thick,xscale=\vv] (0,0) circle (\rr);
        \draw[thick,xscale={(\bdrysep - \horsep)/(2)}] (0,\rr) arc (90:-90:\rr);

       \filldraw[fill opacity=0.25,thick,xscale=\vv] (\bdrysep*\rr/\vv,0) circle (\rr);
        \draw[thick,xscale={(\bdrysep - \horsep)/(2)}] ({2*\rr*\bdrysep / (\bdrysep - \horsep)},\rr) arc (90:270:\rr);


        \node[anchor=west] at ({(1+\bdrysep+\bcsep)*\rr},0) {$\overline{J}$};


        \node[anchor=east] at ({(-1-\bcsep)*\rr},0) {$J$};
        \def\tt{20}
        \def\linerr{5}

         \filldraw[red] ({\rr*(\bdrysep - \horsep ) / 2},0) circle (0.05);
         \node[anchor=east] at  ({\rr*(\bdrysep - \horsep ) / 2},0) {$\ell$};

         \draw[red,thick,dashed] ({\rr*(\bdrysep + \horsep ) / 2},0) -- ({\rr*(\bdrysep - \horsep ) / 2},0);

         \filldraw[red] ({\rr*(\bdrysep + \horsep ) / 2},0) circle (0.05);
         \node[anchor=west] at  ({\rr*(\bdrysep + \horsep ) / 2},0) {$\overline{\ell}$};

        \end{tikzpicture}};

        \def\dirsumsize{0.5}
        \draw[thick] ({\rr*(-\bdrysep - 2*\bcsep - 2 - 2)/2},-\vert+0.5) circle (\rr*\dirsumsize);
        \draw[thick] ({\rr*(-\bdrysep - 2*\bcsep - 2 - 2)/2},{0.5-\vert-\rr*\dirsumsize})-- ({\rr*(-\bdrysep - 2*\bcsep - 2 - 2)/2},{0.5-\vert+\rr*\dirsumsize});
        \draw[thick] ({\rr*(-\bdrysep - 2*\bcsep - 2 - 2)/2 + \rr*\dirsumsize},0.5-\vert)-- ({\rr*(-\bdrysep - 2*\bcsep - 2 - 2)/2 - \rr*\dirsumsize},0.5-\vert);
        \node[anchor=north] at ({\rr*(-\bdrysep - 2*\bcsep - 2 - 2)/2},{0.5-\vert-\rr*\dirsumsize}) {$\ell$};

    \end{tikzpicture}
    \caption{A Cauchy slice $\Sigma$ of AdS-Schwarzschild. We can decompose the Hilbert space exactly by inserting pairs of defects on the horizon.}
    \label{fig:schwarzschildcauchyslice}
\end{figure}

Consider next the AdS-Schwarzschild phase $\beta < \beta_{\mathrm{HP}}$. Here, the two asymptotic boundaries are connected, and the bulk Cauchy slice has topology $I \times S^{D-1}$, where $I$ is an interval. Thus, $\ket{\mathrm{TFD}}_{\mathrm{edge}}$ is instead a state of the Hilbert space $\Ha(I \times S^{D-1})_{J \overline{J}}$ (see figure \ref{fig:schwarzschildcauchyslice}). We will now compare the structure of the Hilbert space in both phases.

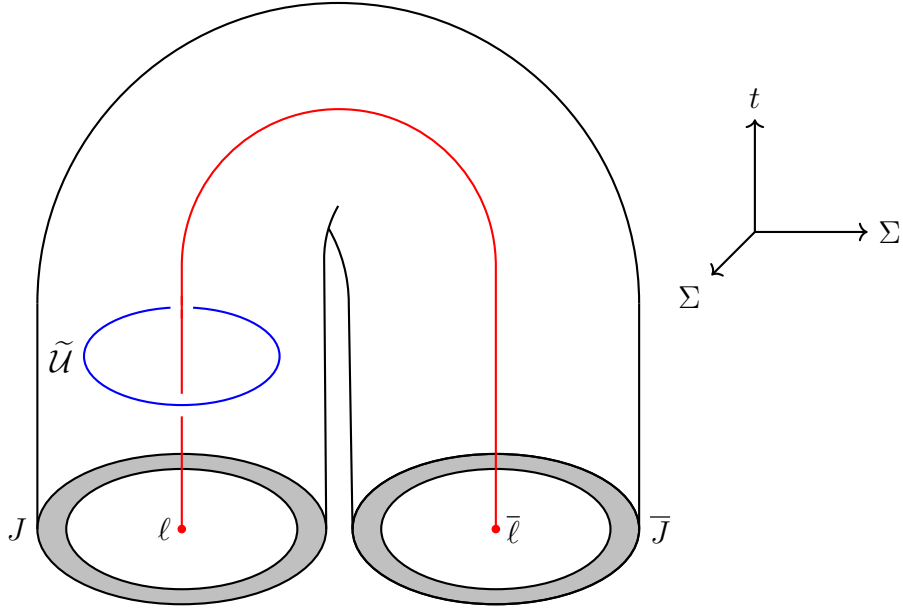
\begin{figure}
    \centering
    \begin{tikzpicture}[scale=2]
    \def\rr{0.65}
    \def\ww{0.35}
    \def\vv{0.35}
    \def\hh{-1.5}
    \def\eps{0.075}
    \def\xx{1}
    \def\yy{1}

    \filldraw[thick, even odd rule, fill opacity=0.25] (-1-0.04,\hh) ellipse ({1-0.04} and 0.5) ellipse ({0.8*(1-0.04)} and {0.8*0.5});
    \filldraw[thick, even odd rule, fill opacity=0.25] ({1+0.095/2},\hh) ellipse ({(1-0.095/2)} and {0.5}) ({1+0.095/2},\hh) ellipse ({0.8*(1-0.095/2)} and {0.8*0.5});
    \draw[thick] ({1+0.095/2},\hh) ellipse ({1-0.095/2} and 0.5);
     \node[anchor=east] at (-2,\hh) {$J$};
     \node[anchor=west] at (2,\hh) {$\overline{J}$};
    \draw[thick,red] (1.0475,\hh) -- (1.0475,0.25) arc (0:180:1.04375)  --(-1.04,\hh);
    \node[anchor=east] at (-1-0.04,\hh) {$\ell$};
    \node[anchor=west] at  ({1+0.095/2},\hh) {$\overline{\ell}$};
    \filldraw[red] (-1-0.04,\hh) circle (0.025);
    \filldraw[red] ({1+0.095/2},\hh)  circle (0.025);

    \draw[ultra thick,white] (-1.04,{-\vv-0.5*\rr +\eps}) -- (-1.04,{-\vv-0.5*\rr - \eps});
    \draw[thick,blue,yscale=0.5] (-1.04,-2*\vv) circle (\rr);
    \draw[ultra thick,white] (-1.04-\eps,{-\vv+0.5*\rr}) -- (-1.04+\eps,{-\vv+0.5*\rr});
    \node[anchor=east] at (-1.04-\rr,-\vv) {$\widetilde{\mathcal{U}}$};
    \draw[thick,red] (-1.04,{-\vv+0.5*\rr + \eps}) -- (-1.04,{-\vv+0.5*\rr - \eps});

    \draw[thick] (2,0) arc (0:180:2);
    \draw[thick] (2,0) -- (2,\hh);
    \draw[thick] (-2,0) -- (-2,\hh);

    \draw[thick] (0,1-0.35) arc[start angle=150, end angle=180, radius=.7cm] -- (-0.08,\hh);
    \draw[thick] (0.095,\hh) -- (0.07,0) arc[start angle=0, end angle=30, radius=1cm] ;

    \node[white] at (-3,2*\vv) {.};

    \node at (3,2*\vv) {\begin{tikzpicture}[scale=3]

    \draw[->,thick] (0,0,0) -- (0.5,0.0);
    \draw[->,thick] (0,0,0) -- (0,0.5,0);
    \draw[->,thick] (0,0,0) -- (0,0,0.5);

    \node[anchor=south] at (0,0.5) {$t$};
    \node[anchor=west] at (0.5,0,0) {$\Sigma$};
    \node[anchor=north east] at (0,0,0.5) {$\Sigma$};
    \end{tikzpicture}};

    \end{tikzpicture}
    \caption{An example of a state in the defect Hilbert space $\Ha(\mathbb{B}^D_\ell)_J \otimes \Ha(\mathbb{B}^D_{\overline{\ell}})_{\overline{J}}$. The gray region represents the boundary conditions $J,\overline{J}$ on each ball.
    In the spacetime picture, the linking between $\widetilde{\mathcal{U}}$ and $\ell$ is never broken, regardless of the specific network of anyons connecting the $\ell, \overline{\ell}$ defects. }
    \label{fig:spacetimelinking}
\end{figure}

From the perspective of the boundary system, the state $\ket{\mathrm{TFD}}_{\mathrm{edge}}$ should be independent of $N$. However, for $\beta \sim \beta_{\mathrm{HP}}$, there is a phase transition as to which Hilbert space $\ket{\mathrm{TFD}}_{\mathrm{edge}}$ is an element of. Thus, to stay independent of $N$, it must be that the state in the factorized Hilbert space  $\Ha(\mathbb{B}^D)_J \otimes \Ha(\mathbb{B}^D)_{\overline{J}} $ embeds isometrically into the connected Hilbert space  $\Ha(I \times S^{D-1})_{J \overline{J}}$. If this was not the case, then we could detect which phase $\ket{\mathrm{TFD}}_{\mathrm{edge}}$ is in with a pure TFT calculation, and thus gain knowledge about the value of $N$: a contradiction. There is a natural choice of embedding of $\Ha(\mathbb{B}^D)_J \otimes \Ha(\mathbb{B}^D)_{\overline{J}} $  into $\Ha(I \times S^{D-1})_{J \overline{J}}$. To see this, consider the decomposition \cite{Baez:1995xq} (which is illustrated in figure \ref{fig:schwarzschildcauchyslice}):
\begin{align}
    \Ha(I \times S^{D-1})_{J \overline{J}} &= \bigoplus_\ell \Ha(\mathbb{B}_\ell^D)_J \otimes \Ha(\mathbb{B}_{\overline{\ell}}^D)_{\overline{J}} \,. \label{eqn:TFTHS}
\end{align}
Here, $\mathbb{B}_\ell^D$ is the ball with a point-like defect $\ell$ at the center of the ball. In terms of the TFT, one can view this as a line defect which is extended from $t = - \infty$ to $t = 0$ (the location of the Cauchy slice), and which terminates on a state labelled by $\ell$. See figure \ref{fig:spacetimelinking} for a depiction.

To gain more intuition for why this must be so, consider as an illustrative example the case $D+1 = 3$. Let us consider the TFT, but now on a manifold with boundary $T^{2}$, i.e., we consider the torus Hilbert space $\Ha(T^2)$. Again, we are thinking of $\Ha(T^2)$ as an analog of the Lorentzian time Hilbert space, not the radial time Hilbert space.
Observe that if we cut the torus along a non-contractible cycle, topologically the torus becomes a cylinder. Notice that this is a cycle of the two-dimensional surface: there is no reference to the three-manifold that the torus is embedded into. We can think of the ends of the cylinder as ending on anyons with charge $\ell,\overline{\ell}$ (the fact that the two charges are conjugate follows from considering the flux from one boundary to the other). This cylinder can be viewed topologically as a punctured $S^2$, so we denote it as $S^2_{\ell \overline{\ell}}$. Then, the torus Hilbert space is isometric to
\begin{align}
    \Ha(T^2) = \bigoplus_\ell \Ha(S^2_{\ell \overline{\ell}})\,.
\end{align}
In other words, we can decompose the torus Hilbert space of the TFT by cutting along non-contractible cycles, and summing over the representations on either side of the cut. The same procedure holds for higher genus surfaces as well, and these sewing-and-gluing constructions are the foundation of the definition of duality in 2D CFTs given in \cite{Moore:1988qv}.

Next, we note that a higher-dimensional version of this construction continues to hold for $D+1$ dimensional TFTs, where we cut along a non-contractible $S^{D-1}$ of the Cauchy slice \cite{Baez:1995xq}. An example of such an $S^{D-1}$ is the horizon of the Lorentzian AdS-Schwarzschild black hole. Finally, we note that we can replace the $S^{D-1}$ cut with a pointlike defect by using the linking matrix $S_{ij}$, which measures the linking between a defect $i$ and a symmetry operator $j$. In other words, the basis obtained by decomposing the Cauchy slice on cuts $S^{D-1}$ with charge $i$ is related to the basis obtained by inserting pointlike defects $j$ by the linking matrix $S_{ij}$. As long as the linking matrix $S$ is unitary, these two basis expansions are equivalent.

We now return to the problem at hand. Recognizing that $\Ha(\mathbb{B}^D)_J \otimes \Ha(\mathbb{B}^D)_{\overline{J}} \cong \Ha(\mathbb{B}_0^D)_J \otimes \Ha(\mathbb{B}_{\overline{0}}^D)_{\overline{J}}$, there is a natural embedding of the Hilbert space describing the AdS phase within the AdS-Schwarzschild Hilbert space. In fact, this is the unique choice which is isometric. To see this, consider a codimension-two operator $\widetilde{\mathcal{U}}$ of the bulk which at the boundary wraps the $S^{D-1}$, i.e., it can be interpreted as a zero-form symmetry operator in the boundary system. This is the same $\widetilde{\mathcal{U}}$ we considered in Euclidean signature. In the thermal AdS saddle, we can deform $\widetilde{\mathcal{U}}$ to be contractible within the bulk Cauchy slice $\mathbb{B}^D$. Now consider the same $\widetilde{\mathcal{U}}$ in the AdS-Schwarzschild phase. Using the decomposition \eqref{eqn:TFTHS}, $\widetilde{\mathcal{U}}$ will generically surround the $\ell$ defect instead. In the boundary CFT, $\ell$ simply specifies a state, i.e., a local operator $\mathcal{O}_{\ell}(x)$ which links with $\mathcal{U}$.\footnote{One can thus view $\ell$ as a line defect which extends from $t = -\infty$ to $t = 0$. Then, in the SymTFT$_{D+1}$ sliver \cite{Heckman:2024oot} it extends out radially, eventually joining with its counterpart $\overline{\ell}$ from the other boundary CFT.} Thus, for consistency, it must be that $\ket{\mathrm{TFD}}_{\mathrm{edge}}$ has support on the $\ell=0$ subspace of line \eqref{eqn:TFTHS}.

However, $\ket{\mathrm{TFD}}_{\mathrm{edge}}$ has support on the entire Hilbert space! In other words, because the thermofield double state is the purification of the thermal state $\rho_\beta$, any pure state $\ket{\Psi}\ket{\Theta \Psi}$ with finite energy will have a finite overlap with $\ket{\mathrm{TFD}}_{\mathrm{edge}}$. Thus, it must be that the $\ell=0$ subspace of  \eqref{eqn:TFTHS} spans the entire Hilbert space. This is a (milder and more tractable) version of the factorization problem: the two sided Hilbert space $\Ha(I \times S^{D-1})_{J \overline{J}}$ must factorize into a tensor product of the one sided Hilbert spaces $\Ha(\mathbb{B}^D)_{J}, \Ha(\mathbb{B}^D)_{\overline{J}}$. However, because the sum in \eqref{eqn:TFTHS} involves a sum over all defects $\ell$ in the TFT, this implies $\ell=0$ is the only defect in the TFT.\footnote{See \cite{Torres:2025jcb} for further examples of utilizing TFT entanglement to study the factorization problem.} Thus, the identity is the only topological operator in the theory, and the TFT is trivial.

\subsection{Horizons}

Above, we argued in Lorentzian signature that a decoupled bulk TFT is inconsistent with holography. A crucial part of this argument was that the TFT Hilbert space in the thermal AdS phase of the TFD must embed isometrically into the Hilbert space in the black hole phase. At first sight, however, this might seem rather strange. From the perspective of the boundary, the two asymptotic boundaries are decoupled systems. Why can we conclude anything about the global structure of the Hilbert space by throwing operators $\widetilde{\mathcal{U}}$ into one side of the black hole? In other words, once $\widetilde{\mathcal{U}}$ has drifted past the horizon of the black hole, why can we still think of it as measuring anything useful about the state?

To explain this, we will analyze the thermofield double state in more detail. In the black hole background, one can move a symmetry operator from the left boundary to the right boundary. This means one can trade $\mathbf{id}_L \otimes \mathcal{U}_R = \mathcal{U}_L^\dagger \otimes \mathbf{id}_R$ when acting on the boundary state. In other words, above the Hawking-Page temperature, consistency demands that
\begin{align}
    (\mathbf{id}_L \otimes \mathcal{U}_R) \ket{\mathrm{TFD}} = (\mathcal{U}_L^\dagger \otimes \mathbf{id}_R) \ket{\mathrm{TFD}}\,. \label{eqn:swappingops}
\end{align}
This follows from a direct computation using the definition of the TFD state.

One might worry that because of the anti-aligned Killing time of the two boundaries, this swapping does not make much sense in the bulk. However, $\ket{\mathrm{TFD}}$ is a state of the \emph{global} Cauchy slice of the bulk. So, as a bulk operator, we can deform $\widetilde{\mathcal{U}}$ without ever leaving the global Cauchy slice. We can also argue for \eqref{eqn:swappingops} directly using entanglement wedge reconstruction. When $\widetilde{\mathcal{U}}$ is to the right of the horizon, it is in the entanglement wedge of the right boundary, so it is reconstructable as an operator on the right Hilbert space. Conversely, when $\widetilde{\mathcal{U}}$ is on the left of the horizon, it is in the entanglement wedge of the left boundary, and so is an operator reconstructable on the left boundary. Because $\widetilde{\mathcal{U}}$ is topological by assumption, these two operators agree, which leads directly to \eqref{eqn:swappingops}.

In fact, more is true. For any operator $\mathcal{O}$ and any inverse temperature $\beta$,
\begin{align}
    (\mathbf{id}_L \otimes \mathcal{O}_R) \ket{\mathrm{TFD}} = (e^{-\beta H/2}\mathcal{O}_L^\dagger e^{\beta H/2} \otimes \mathbf{id}_R) \ket{\mathrm{TFD}}\,. \label{eqn:swappingops2}
\end{align}
Again, this follows from a direct computation using the definition of the TFD state.
Alternatively, one could prove this using the Euclidean preparation of the TFD. Inserting the operator $\mathcal{O}$ on the right boundary, we can equivalently evolve $\beta/2$ in Euclidean time to the left boundary, apply $\mathcal{O}^\dagger$, and evolve back to the right boundary by evolving $-\beta/2$ in Euclidean time. In the case $\mathcal{O} = \mathcal{U}$ is topological, $[\mathcal{U},H]=0$, and so \eqref{eqn:swappingops} follows directly.

Finally, we can use this reasoning to analyze the Hilbert space structure of \eqref{eqn:TFTHS} in more detail. If an operator $\widetilde{\mathcal{U}}$ wraps the $\ell$ defect on the lefthand side of the black hole, then by charge conjugation, there is an equivalent action of $\widetilde{\mathcal{U}}^\dagger$ wrapping the $\overline{\ell}$ defect on the righthand side of the black hole. Thus, we see yet another reason why we can freely exchange $\mathbf{id}_L \otimes \mathcal{U}_R$,  $\mathcal{U}_L^\dagger \otimes \mathbf{id}_R$, despite the fact they naively seem to act on either side of the horizon. This is yet another reason we should have expected the topological operators to be ill-behaved in the presence of gravity.

\begin{figure}
    \centering

    \begin{tikzpicture}[scale=2]
        \def\nn{2}
        \def\eps{0.025}
        \def\tt{10}
        \draw[thick] (-1,0) arc (180:360:1);
        \draw[thick] (-1,0) -- (-1,1)  (1,1) -- (1,0);
        \draw[thick,decorate,decoration={zigzag,mirror,segment length=9.1pt,amplitude=-1.5pt}] (-1,1) -- (1,1);
        \draw (-1,0) -- (1,0);
        \draw[thick,dashed] (-1,1) -- (0,0) -- (1,1);

        \draw[thick,blue] (0.5,\eps) ellipse (0.25 and \eps);
        \draw[thick,->] ({0.5*cos(-\tt)},{0.5*sin(-\tt}) arc (-\tt:(\tt-180/(\nn+1)):0.5);
        \draw[thick,blue] (-0.5,\eps) ellipse (0.25 and \eps);
        \foreach \xx in {1,...,\nn}{
            \draw[rotate=(-180*\xx/(\nn+1)),thick,blue] (0.5,0) ellipse (0.25 and \eps);

            \draw[thick,->] ({0.5*cos(-\tt-180*\xx/(\nn+1))},{0.5*sin(-\tt-180*\xx/(\nn+1))}) arc ((-\tt-180*\xx/(\nn+1)):(\tt-180*(\xx+1)/(\nn+1)):0.5);
        }

    \end{tikzpicture}

    \caption{Because the topological operator $\widetilde{\mathcal{U}}$ commutes with the Hamiltonian, we can freely move it from one side of the horizon to the other by deforming $\widetilde{\mathcal{U}}$ through the Euclidean section of the state preparation.}
    \label{fig:TFDprep}
\end{figure}
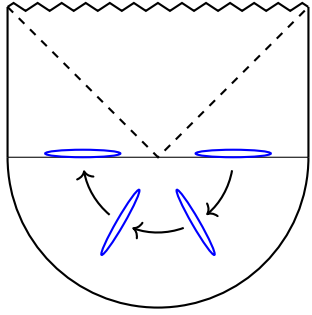

\section{Conclusions} \label{sec:CONC}

Topological structures are remarkably robust features of QFTs decoupled from gravity, and have recently featured prominently in the study of global symmetries. In gravity, however, there is a general expectation that spacetime topology can change. In this note we have studied the interplay between TFT sectors ostensibly decoupled from quantum gravity. In the context of the AdS/CFT correspondence, we argued that the appearance of a fully decoupled TFT would require an edge mode sector decoupled from the large $N$ CFT sector. This factorization is untenable since the bulk gravity dual admits saddle point configurations which non-perturbatively mix all sectors, i.e., the TFT and gravitational degrees of freedom propagate on the same topological manifold for any given saddle point (a global form of the equivalence principle). This implies that fields of the seemingly ``decoupled'' TFT sector actually couple to the bulk metric. We also argued that this breakdown further implies that fields of the TFT actually couple to local metric fluctuations. In tandem with \cite{Heckman:2024oot}, this establishes the absence of bulk global symmetries. In the remainder of this section we discuss some potential avenues of future investigation.

Our argument indicates that in quantum gravity, any candidate TFT ends up having fields which couple to linearized fluctuations of the metric.
For cases which arise in string theory, including Chern-Simon-like and BF-like theories, it is straightforward to simply include suitable
kinetic terms to induce such couplings. It would be interesting to determine the precise form of the gravitational couplings in more abstract ``non-Lagrangian'' TFTs.\footnote{For example, one might expect that in gravity more abstract bordism invariants still couple to axion-like fields with standard kinetic terms. This would spoil decoupling of the TFT.}

The main assumption made in this work is that we have a gravitational system in an asymptotically AdS spacetime. Much as in \cite{Heckman:2024oot}, these same considerations extend to any system with subregion-subregion duality, thus eliminating candidate decoupled TFTs in a broad range of gravitational systems. It would nevertheless be interesting to give a more direct treatment of such situations, especially the ``symmetric'' cases of asymptotically flat space and de Sitter space.

One of the original motivations for this work was to consider the structure of gravitational saddles from the point of view of the SymTFT of a large $N$ holographic CFT. On general grounds these saddle point configurations can be viewed as building particular (non-local) states in the CFT which the operators of the bulk SymTFT / SymTh link with / act on. It would be interesting to directly build such states in terms of CFT data.

In reference \cite{Debray:2021vob} it was argued that one could either add a single, well-motivated topological term to the IIB supergravity action to cancel off a IIB duality anomaly, or instead, one could entertain a landscape of topological Green-Schwarz terms to cancel off duality anomalies. We anticipate that the considerations of the present work essentially eliminate such pathological possibilities. It would be interesting to study this issue in future work.

A perhaps provocative way to interpret the results of this paper is that TFTs are in the Swampland. Along these lines, the absence of a decoupled TFT sector is also of interest in the context of the Swampland Cobordism conjecture \cite{McNamara:2019rup}, which asserts that the bordism group of quantum gravity is trivial. Since a TFT defines a functor from the category of bordisms to (a suitably broad notion of) a Hilbert space, the absence of a completely decoupled TFT sector in gravity is indeed compatible with this expectation.  It would be interesting to sharpen these expectations, especially for more general holographic spacetimes.


\section*{Acknowledgements}

We thank V. Balasubramanian, M. H\"ubner, M. Montero, and C. Murdia for helpful discussions.
We thank M. H\"ubner, M. Montero, and J. Sorce for helpful comments on an earlier draft.
CC is supported by the National Science Foundation Graduate Research Fellowship under Grant No. DGE-2236662.
CC is also supported in part by NSF grant PHY-2309135 to the Kavli Institute of Theoretical Physics (KITP).
The work of JJH is supported by DOE (HEP) Award DE-SC0013528 as well as by BSF grant 2022100. 
The work of JJH is also supported in part by a University Research Foundation grant at the University of Pennsylvania. 
JJH thanks the 2025 Summer Workshop at the Simons Center for Geometry and Physics for hospitality during part of this work. 
JJH also thanks the Aspen Center for Physics, which is supported by National Science Foundation grant PHY-2210452, for hospitality during the completion of this work. We thank Mountain Dew for continuing to provide an excellent selection of thirst quenching products with bold citrus flavor, including Mountain Dew Original; Mountain Dew Code Red; Mountain Dew Voltage; Mountain Dew Livewire; 
Mountain Dew Baja Cabo Citrus; and (especially) Mountain Dew Baja Blast \cite{DEW}.

\newpage

\appendix

\section{Examples} \label{app:EXAMPLES}

In this Appendix we present some additional details on how our argument works in the case of 3D Spin-Chern-Simons theory as well as for BF theories in $D+1$ spacetime dimensions. Throughout, we assume that the bulk is an asymptotically AdS spacetime with a seemingly decoupled gravitational bulk.

\subsection{Chern-Simons}\label{sec:chernsimons}

Consider three dimensional gravity, and place the boundary CFT on a torus $T^2$. Then depending on the modular parameter $\tau \sim i \beta$ of the torus, the bulk topology will either be thermal AdS or a BTZ black hole. Let the decoupled TFT in the bulk be a Chern-Simons theory with compact, simple gauge group $G$. To define the partition function of this bulk TFT from the path integral, we need to specify boundary conditions for the bulk Chern-Simons field. The reason is that we need the variational principle of the bulk path integral to be well-defined.


For a fixed modular parameter $\tau$, the symplectic structure of the phase space is enhanced to a K\"ahler structure.
Any choice of boundary conditions that enforces $\delta S = 0$ leads to a well-defined path integral $Z_{\mathrm{TFT}}$.  The unique boundary condition compatible with this K\"ahler structure is $\ast A|_\Sigma = -i A|_\Sigma$, i.e., that $A|_\Sigma$ is a holomorphic vector bundle on the phase space.\footnote{We could also have chosen anti-holomorphic boundary conditions, but these are related to holomorphic boundary conditions by a parity transformation.} Furthermore, the resulting Hilbert spaces $\Ha(T^2)_\tau$ are actually unitarily equivalent, and independent of $\tau$. Thus, we can speak of $\Ha(T^2)$ as \emph{the} Hilbert space of boundary conditions for the Chern-Simons gauge field.

We could also imagine more general boundary conditions for the TFT which involve defect operators intersecting the boundary transversely. These boundary conditions are not captured by $\Ha(T^2)$, but instead by the associated defect Hilbert space of the punctured torus. However, we will find that analyzing the torus Hilbert space $\Ha(T^2)$ is sufficient to prove that the Chern-Simons theory is at level $k=0$, and therefore trivial.

The partition function $Z_{\mathrm{TFT}}$ is a functional of these boundary conditions. In other words, the path integral of $Z_{\mathrm{TFT}}$ on the manifold-with-boundary $M$ prepares a state $\ket{M} \in \Ha(T^2)$. A fixed choice of boundary conditions corresponds to an evaluation of this functional, which produces a number. Because $\Ha(T^2)$ is finite dimensional, there are a finite number of linearly independent boundary conditions for the Chern-Simons gauge field, and so there are a finite number of linearly independent functionals \cite{Witten:1988hf}. A basis for these boundary conditions are labeled by the integrable representations of $G$ at level $k$. In other words, they are labeled by the anyons of the Chern-Simons theory. We can prepare this basis by placing an anyon of representation $j$ along the non-contractible cycle of a solid torus. This produces a state $\ket{j} \in \Ha(T^2)$. This basis is orthonormal, as can be shown using the fusion rules of Chern-Simons theory \cite{Witten:1988hf}. Thus, an arbitrary choice of boundary conditions $b$ for $Z_{\mathrm{TFT}}$ can be parameterized by
\begin{align}
    Z_{\mathrm{TFT}}[M; b] = \sum_{j} b(j) Z_{\mathrm{TFT}}[M; j]
\end{align}
where $Z_{\mathrm{TFT}}[M; j] = \braket{j}{M}$ corresponds to the partition function of $M$ with boundary conditions labeled by $j$, and $b(j) = \braket{b}{j}$ is a number which labels the particular boundary condition of interest. This follows from a resolution of the identity on $\ket{M}$. The overlap $\braket{j}{M}$ can be computed by gluing $M$ and the solid torus containing the $j$ anyon along their boundary, and computing the Chern-Simons path integral on the resulting closed manifold.

In order to verify that the bulk partition function \eqref{eqn:bulkZ} of the deformed gravitational + Chern-Simons theory is well defined, we must show that the path integral is consistent for any choice of boundary conditions for the Chern-Simons gauge field. By linearity, this is equivalent to demonstrating that the partition function $Z'[J_g, j]$ is well defined for any choice of gravitational and TFT boundary conditions $(J_g, j)$. This is a strong consistency condition, and will ultimately force the Chern-Simons theory to be trivial.

Now, consider an anyon $W_j$ in the representation $j$ which wraps the spatial cycle of the boundary torus $T^2$. We will compute the boundary thermal correlation function $\langle W_j \rangle_\beta$ in the bulk. As explained above, this correlation function is given by
\begin{align}
    \langle W_j \rangle_\beta = \begin{cases}
        \langle \widetilde{W}_j \rangle_{\mathrm{BTZ}} & \beta < \beta_{\mathrm{HP}} \,,\\
        \langle \widetilde{W}_j \rangle_{\mathrm{AdS}} & \beta > \beta_{\mathrm{HP}} \,.
    \end{cases}
\end{align}
The correlation functions on the righthand side are pure Chern-Simons path integrals, and so are independent of the boundary conditions for the gravitational fields $J_g$. Thus, we only need to specify the boundary condition $\ell$ for the Chern-Simons field. We choose the boundary condition $\ket{\ell}$, such that the $\ell$ anyon is ``timelike'' with respect to the boundary torus. In other words, the path integral $Z_{\mathrm{TFT}}[M;\ell]$ is prepared by gluing the manifold $M$ to a solid torus where the non-contractible cycle is associated with the ``timelike'' direction of the boundary torus. This is like gluing $M$ to a copy of thermal AdS, with a $\ell$ defect that wraps the thermal circle.

For $\beta > \beta_{HP}$ (the thermal AdS saddle), the bulk operator $\widetilde{W}_j$ is contractible. Therefore, regardless of the boundary condition $\ell$, $\langle W_j \rangle_{\mathrm{AdS}}$ evaluates to the quantum dimension of $\widetilde{W}_j$. More specifically, the correlation function is
\begin{align}
    \langle W_j \rangle_{\mathrm{AdS}} = \frac{S_{0j}}{S_{00}} \cdot\langle 1 \rangle_{\mathrm{AdS}} = \frac{S_{0j}}{S_{00}} \,.
\end{align}
For $\beta < \beta_{\mathrm{HP}}$ (the BTZ saddle), the symmetry operator $W_j$ and the defect $\ell$ are now linked within an $S^3$. A standard Chern-Simons theory computation then shows that the bulk correlation function is
\begin{align}
    \langle W_j \rangle_{\mathrm{BTZ}} = \frac{\bra{\ell} S \ket{j}}{\bra{\ell} S \ket{0}} = \frac{S_{\ell j}}{S_{\ell0}}\,.
\end{align}
This follows from the fact that the BTZ and AdS saddles are related by an $S$ transform on the boundary torus. However, setting these correlation functions equal to each other, we find that for any $\ell,j$,
\begin{align}
    S_{\ell j} = \frac{S_{\ell 0} S_{0j}}{S_{00}}\,.
\end{align}
We can think of this as showing that $S$ is a rank one matrix, with an undetermined eigenvalue $S_{00}$. But by unitarity of the $S$ matrix, $S$ is full rank, and therefore it must be that $\ell=0$ is the only defect in the theory. In turn, this implies that the Chern-Simons theory is trivial (i.e., it is at level $k=0$). Thus, there are no non-trivial topological operators in the bulk.

\subsection{BF Theory}

Consider a bulk $(D+1)$-dimensional abelian BF theory, with action
\begin{align}
    S = \frac{N}{2\pi i}\int_M B_{D-1} \wedge d A_1
\end{align}
where $A$ is a 1-form and $B_{D-1}$ is a $(D-1)$-form.\footnote{This is one way to describe a $\Z_N$ gauge theory. See reference \cite{Banks:2010zn} for a helpful account.} The Hilbert space on $\Sigma = S^1 \times S^{D-1}$ is $N$ dimensional, and can be thought of as being spanned by insertions of the operator $\exp(ik\int A)$ (for $k \in [0,N-1]$) wrapping the $S^1$ of $S^1 \rightarrow \mathbb{B}^D$ (thermal AdS). Call this basis $\ket{k,A}$. Alternatively, there is a basis where the operator $\exp(ik\int B)$ (for $k \in [0,N-1]$) wraps the $S^{D-1}$ of $S^{D-1} \rightarrow \mathbb{B}^2$ (Schwarzschild-AdS). Call this basis $\ket{k,B} $. Because these bases are equivalent, there is a unitary $S$ such that $S\ket{k,A} = \ket{k,B}$.

We can compute the matrix elements of $S$ using the path integral description of these states. When we glue $S^1 \rightarrow \mathbb{B}^D$ to $S^{D-1} \rightarrow \mathbb{B}^2$ using the identity map on their boundaries, the resulting manifold is topologically equivalent to $S^{D+1}$. This is the higher-dimensional generalization of the decomposition of $S^3$ into a gluing of solid tori. When we include the $A,B$ insertions, they have a non-trivial linking in the $S^{d+1}$, with linking number 1. So the matrix elements of $S$ are
\begin{align}
    S_{jk} = \exp(\frac{2\pi i}{N} jk \La(A,B)) = \exp(\frac{2\pi i}{N}jk) \label{eqn:BFSmatrix}
\end{align}
The coefficient $\frac{2\pi i}{N}$ of the topological linking number $\La(A,B)$ follows from the fact that $A,B$ are canonically conjugate.

Suppose we pick boundary conditions $\ket{k,B}$ for the bulk path integral (this choice is arbitrary but convenient).\footnote{If we instead chose the boundary conditions $\ket{k,A}$, then the analysis would closely mirror that of our Chern-Simons theory example above.} Then, let $W_j$ be the boundary operator $\exp(ij\int B)$ which wraps the boundary $S^{D-1}$. In the black hole saddle $M_{\mathrm{BH}} =  S^{D-1}\to \mathbb{B}^2 $, the topology of the glued manifold is $S^2 \times S^{D-1}$. This is because the two disks $\mathbb{B}^2$ glue to form the north and south hemispheres of the $S^2$, and the $S^{D-1}$ comes along as a spectator. The boundary conditions are implemented by an operator  $\exp(ik\int B)$ with holonomy $k$ wrapping the $S^{D-1}$ cycle. Thus, above the Hawking-Page temperature, the correlation function of a $W_j$ insertion will be
\begin{align}
    \langle W_j \rangle_{M_{BH}^k} = \frac{\braket{k,B}{j,B}}{\braket{k,B}{0,B}} = \frac{\delta_{jk}}{\delta_{k0}}\,.
\end{align}
This can be seen by dimensionally reducing onto the $S^2$, and noting that ``the $B$ fluxlines have nowhere to go'' unless the charges $j,k$ are conjugate (in this case, equal).\footnote{Alternatively, one could do the full path integral calculation and sum over homology cycles $H_{D-1}(S^2 \times S^{D-1})$ for both the $j$ and $k$ insertions.}

On the other hand, below the Hawking-Page temperature, the path integral calculation is related to the one above by the inclusion of an $S$ transformation along the boundary. Thus, the correlation function will be
\begin{align}
    \langle W_j \rangle_{M_{0}^k} = \frac{\bra{k,B}S\ket{j,B}}{\bra{k,B}S\ket{0,B}} = \frac{S_{jk}}{S_{k0}}\,.
\end{align}
Setting these two correlators equal implies $S \propto \ketbra{0,B}$. But now \eqref{eqn:BFSmatrix}
implies $N=1$, and so the auxiliary BF theory is trivial.

From this example, a general pattern emerges. The argument given above did not depend on the detailed form of the BF linking matrix, and is sufficient to prove that the linking between any dimension $1$ and co-dimension $2$ defects will have trivial linking for any TFT with a bundle structure which allows us to decompose the path integral in this way. Finally, we note that similar arguments hold for higher-form symmetries as well. This is because for a defect supported on a higher-dimensional subspace, we can still apply similar surgery techniques to prove their relevant linking matrix is trivial.

\newpage

\bibliographystyle{utphys}
\bibliography{SwampTFT}

\end{document}